\documentclass[iop]{emulateapj}% change onecolumn to iop for a fancier appearance

\let\pwiflocal=\iffalse \let\pwifjournal=\iffalse

\usepackage{amsmath,amssymb}
\usepackage[breaklinks,colorlinks,urlcolor=blue,citecolor=blue,linkcolor=blue]{hyperref}
\usepackage{epsfig}    
\usepackage{graphicx}    
\usepackage{lineno}
\usepackage{natbib}
\usepackage{bigints}
\usepackage{epstopdf}
\newcommand{\angstrom}{\mbox{\normalfont\AA}}

\usepackage[T1]{fontenc}
\pwifjournal\else
  \usepackage{microtype}
\fi

\pwifjournal\else
  \makeatletter
  \renewcommand\plotone[1]{%
    \centering \leavevmode \setlength{\plot@width}{0.95\linewidth}
    \includegraphics[width={\eps@scaling\plot@width}]{#1}%
  }%
  \makeatother
\fi

\makeatletter

\newcommand\@simpfx{http://simbad.u-strasbg.fr/simbad/sim-id?Ident=}

\newcommand\MakeObj[4][\@empty]{% [shortname]{ident}{url-escaped}{formalname}
  \pwifjournal%
    \expandafter\newcommand\csname pkgwobj@c@#2\endcsname[1]{\protect\object[#4]{##1}}%
  \else%
    \expandafter\newcommand\csname pkgwobj@c@#2\endcsname[1]{\href{\@simpfx #3}{##1}}%
  \fi%
  \expandafter\newcommand\csname pkgwobj@f#2\endcsname{#4}%
  \ifx\@empty#1%
    \expandafter\newcommand\csname pkgwobj@s#2\endcsname{#4}%
  \else%
    \expandafter\newcommand\csname pkgwobj@s#2\endcsname{#1}%
  \fi}%

\newcommand\MakeTrunc[2]{% {ident}{truncname}
  \expandafter\newcommand\csname pkgwobj@t#1\endcsname{#2}}%

\newcommand{\obj}[1]{%
  \expandafter\ifx\csname pkgwobj@c@#1\endcsname\relax%
    \textbf{[unknown object!]}%
  \else%
    \csname pkgwobj@c@#1\endcsname{\csname pkgwobj@s#1\endcsname}%
  \fi}
\newcommand{\objf}[1]{%
  \expandafter\ifx\csname pkgwobj@c@#1\endcsname\relax%
    \textbf{[unknown object!]}%
  \else%
    \csname pkgwobj@c@#1\endcsname{\csname pkgwobj@f#1\endcsname}%
  \fi}
\newcommand{\objt}[1]{%
  \expandafter\ifx\csname pkgwobj@c@#1\endcsname\relax%
    \textbf{[unknown object!]}%
  \else%
    \csname pkgwobj@c@#1\endcsname{\csname pkgwobj@t#1\endcsname}%
  \fi}

\makeatother

\pwifjournal\else
  \usepackage{etoolbox}
  \makeatletter
  \patchcmd{\NAT@citex}
    {\@citea\NAT@hyper@{%
       \NAT@nmfmt{\NAT@nm}%
       \hyper@natlinkbreak{\NAT@aysep\NAT@spacechar}{\@citeb\@extra@b@citeb}%
       \NAT@date}}
    {\@citea\NAT@nmfmt{\NAT@nm}%
     \NAT@aysep\NAT@spacechar\NAT@hyper@{\NAT@date}}{}{}
  \patchcmd{\NAT@citex}
    {\@citea\NAT@hyper@{%
       \NAT@nmfmt{\NAT@nm}%
       \hyper@natlinkbreak{\NAT@spacechar\NAT@@open\if*#1*\else#1\NAT@spacechar\fi}%
         {\@citeb\@extra@b@citeb}%
       \NAT@date}}
    {\@citea\NAT@nmfmt{\NAT@nm}%
     \NAT@spacechar\NAT@@open\if*#1*\else#1\NAT@spacechar\fi\NAT@hyper@{\NAT@date}}
    {}{}
  \makeatother
\fi

\newcommand{\um}{$\mu$m}
\newcommand{\fbol}{$F_{\mathrm{bol}}$}

\newcommand{\hipp}{{\it Hipparcos}}

\newcommand\teff{\ensuremath{T_\text{eff}}}

\newcommand\weff{W$_{\rm{eff}}$}

\providecommand{\adsurl}[1]{\href{#1}{ADS}}

\slugcomment{Accepted to PASP}

\shorttitle{Revised zero points and system profiles}
\shortauthors{Mann \& von Braun}

\begin{document}

\title{Revised Filter Profiles and Zero Points for Broadband Photometry}

\author{Andrew W. Mann\altaffilmark{1,2}, Kaspar von Braun\altaffilmark{3,4}}
  
\altaffiltext{1}{Harlan J. Smith Fellow, Department of Astronomy, The University of Texas at Austin, Austin,
TX 78712, USA} 
\altaffiltext{2}{Visiting Researcher, Institute for Astrophysical Research, Boston University}
\altaffiltext{3}{Lowell Observatory, 1400 W. Mars Hill Rd., Flagstaff, AZ, USA}
\altaffiltext{4}{Max-Planck-Institute for Astronomy (MPIA), Konigstuhl 17, 69117 Heidelberg, Germany}

\begin{abstract}
Estimating accurate bolometric fluxes for stars requires reliable photometry to absolutely flux calibrate the spectra. This is a significant problem for studies of very bright stars, which are generally saturated in modern photometric surveys. Instead we must rely on photometry with less precise calibration. We utilize precisely flux-calibrated spectra to derive improved filter bandpasses and zero points for the most common sources of photometry for bright stars. In total we test 39 different filters in the General Catalog of Photometric Data as well as those from Tycho-2 and {\it Hipparcos}. We show that utilizing inaccurate filter profiles from the literature can create significant color terms resulting in fluxes that deviate by $\gtrsim$10\% from actual values. To remedy this we employ an empirical approach; we iteratively adjust the literature filter profile and zero point, convolve it with catalog spectra, and compare to the corresponding flux from the photometry. We adopt the passband values that produces the best agreement between photometry and spectroscopy and is independent of stellar color. We find that while most zero points change by $<5\%$, a few systems change by 10--15\%. Our final profiles and zero points are similar to recent estimates from the literature. Based on determinations of systematic errors in our selected spectroscopic libraries, we estimate that most of our improved zero points are accurate to 0.5--1\% or better.  
\end{abstract}


\section{Introduction}\label{sec:intro}
Stellar effective temperatures (\teff) are generally determined through a variety of indirect methods, such as the infrared flux method \citep[IRFM, ][]{Blackwell1977, 2006MNRAS.373...13C, Casagrande2010}, fitting observations to synthetic spectra \citep[e.g.,][]{Valenti:1996}, or using equivalent widths and/or line ratios \citep[e.g.,][]{1973PhDT.......180S, Sousa2010, Sousa2011}. Because of model errors and the need for a zero point for IRFM these methods must be calibrated using direct methods. The principal direct method is interferometry, which can be used to determine a star's angular diameter ($\theta$). With $\theta$ and a measurement of the star's bolometric flux (\fbol) we can calculate effective temperature from a rewritten form of the Stefan-Boltzmann Law:
\begin{equation}\label{eqn:stefan}
\mathrm{T}_{\mathrm{eff}} = \left( \frac{{F}_{\mathrm{bol}}}{\sigma (\theta/2)^2}\right) ^{1/4},
\end{equation}
where $\sigma$ is the Boltzmann constant. Because Equation~\ref{eqn:stefan} defines \teff, it provides an empirical, and accurate method to establish the true \teff\ scale of stars. 

Typically \fbol{} is measured using flux-calibrated empirical templates \citep[e.g.,][]{1998PASP..110..863P} or models \citep{Allard:2012fk}. Broadband photometry is used to inform which template or model is the best match to the star and used to derive the absolute flux calibration \citep[for a more detailed description of this method see][]{van-Belle:2008lr}. Even when using the star's true spectrum instead of a template, the spectrum must be absolutely calibrated using photometry because of slit losses, atmospheric variability, etc.. Thus both methods are sensitive to errors in photometry, particularly errors common to entire photometric systems such as inaccuracies in the zero point and filter passband. Such errors could explain much of the systematic difference seen between \fbol\ values for the same stars when using different photometry \citep{Boyajian2012, 2013ApJ...779..188M}.

Most modern astronomical work can make use of precisely calibrated photometry from surveys such as the Sloan Digital Sky Survey \citep{Stoughton:2002,2009ApJS..182..543A} and the Two-Micron All-Sky Survey \citep[2MASS, ][]{Skrutskie:2006lr}. However, long-baseline interferometry is currently limited to the largest and/or nearest stars, whose photometry is almost always saturated in such surveys. Instead, we must make use of older photometry, often from the General Catalog of Photometric Data\footnote{\href{http://obswww.unige.ch/gcpd/}{http://obswww.unige.ch/gcpd/}} \citep[GCPD, ][]{Mermilliod:1997qe}, which is a collection of photometry on $>200,000$ stars from $>2500$ papers. The GCPD contains more than 80 systems over a wide range of wavelengths, which makes it particularly useful for flux calibrating spectra and deriving spectral-energy distributions for bright stars. However, most of these photometric systems are calibrated no more precisely than 5\% \citep[see][and references within]{Bohlin2014}. 

Recent improvements in reduction techniques and instrumentation for long-baseline optical interferometry \citep[e.g.,][]{Ireland2008, ten-Brummelaar2012} have enabled the measurement of the stellar angular diameters ($\theta$) to better than $2\%$ for stars of sufficient brightness and angular size \citep[e.g., ][]{Berger2006, 2013ApJ...771...40B,2013ApJ...767..127H, 2014MNRAS.438.2413V}. Errors of 1--2\% in the angular diameter result in errors in \teff{} of just 0.5--1\% ($\simeq$30~K for a Sun-like star) ignoring errors in \fbol. In order to be limited by the error in $\theta$ (which is typically the harder quantity to measure) \fbol{} must be measured with an error $\lesssim2\%$. More importantly, \fbol{} values need to be free of systematic errors, which could conceivably arise from zero point and filter profile errors in the photometry, deviations in the shape of the spectrum, or inaccurate corrections for interstellar reddening.

When converting magnitudes into fluxes, a slightly erroneous filter profile (i.e., different from the true system profile) or red leak will result in color-terms \citep{1990PASP..102.1181B, Bessell:1988qy}. Specifically, flux derived from convolving the filter profile with the spectrum will differ from the flux calculated from corresponding photometry as a function of the color of the star. One can reverse-engineer the filter passband by calculating synthetic fluxes from precisely calibrated spectra, then comparing these values to the corresponding flux derived from the photometry of the same stars \citep[e.g., ][]{1996BaltA...5..459S, 2006AJ....131.1184M}. The filter profile can be adjusted until the difference between the synthetic and observed fluxes is a constant independent of the star's color, and then the zero point adjusted until that constant is on average $\simeq0$. This method is entirely empirical, and takes into account the full system throughput, but is limited by systematic errors in the spectral flux calibration. 

In this paper we make use of several well-calibrated spectroscopic libraries to verify and derive revised filter bandpasses, zero points, and corresponding errors for photometry of bright stars (GCPD, Tycho-2, and \hipp). In Section~\ref{sec:spsample} we describe our sample of spectroscopic libraries and in Section~\ref{sec:phsample} we give a rundown of the photometric sample. In Section~\ref{sec:analysis} we explain how we derive new filter bandpasses by forcing the synthetic and observed photometry to give consistent fluxes. In Section~\ref{sec:results} we detail our revised properties for different photometric systems, compare our parameters to those in the literature, and estimate the error on our zero points. We conclude in Section~\ref{sec:conclusion} with a brief summary and highlight how our findings can significantly alter conclusions about \fbol{} values for cool stars.

\section{Spectroscopic Sample}\label{sec:spsample}
A spectroscopic sample that is optimal for our purposes would include a large number ($>100$) of stars with photometry in the GCPD covering a wide range in $B-V$ color, and contain spectra with significant wavelength coverage to include as many bands as possible. The MILES library \citep{Sanchez2006, Falcon2011} is one of the largest ($\simeq$1000 stars) and spans a wide range of $B-V$ colors. However, MILES spectra cover only 3525--7000\AA, which excludes sections of the bluest (e.g., Johnson $U$, Geneva $U$, Str\"omgren $u$) and reddest (e.g., Cousins $I$) optical filters. We resolve not to use MILES even for the passbands covered to maintain data homogeneity. Libraries like the Indo-U.S. Library \citep{2004ApJS..152..251V} have both wide wavelength coverage and a large number of stars, but the narrow slit, observing conditions, and joining of multiple observations with different gratings means that the flux calibration is unlikely to have sufficient precision and accuracy for our purposes. 

For optical passbands we use the Next Generation Spectral Library \citep[NGSL,][]{Gregg2006,Heap2007}, which includes spectra of 374 stars spanning $-0.3<B-V<1.8$. NGSL spectra are taken with the Space Telescope Imaging Spectrograph (STIS) on the Hubble Space Telescope (HST) using three different gratings (G230LB, G430L, and G750L). This provides spectral coverage from 2000\AA\ to 10000\AA. The only optical filter in our sample (Section~\ref{sec:phsample}) not covered by NGSL spectra is Johnson $I$. Although the NGSL library contains fewer stars than MILES, the superior wavelength coverage is critical because it enables us to include far more filter systems in our analysis. Further, the relative flux calibration of HST STIS spectra is expected to be accurate to 0.5\% \citep{Bohlin2001,Bohlin:2004, 2004AJ....127.3508B}, which is superior to what can currently be achieved from the ground.

For the NIR bands (Johnson $JHKL$) we utilize the Infrared Telescope Facility (IRTF) spectral library \citep{Cushing:2005lr, Rayner:2009kx}, which includes 210 F, G, K, and M stars with wavelength coverage 8000-50,000\AA. Only $\simeq$ one third of these stars have $JHK$ photometry in the GCPD, but this is better than comparable NIR catalogs, which tend to cover fainter stars that are less likely to be in the GCPD. The catalog also lacks the bluest stars ($B-V<0.15$), but goes significantly redder ($B-V>2$) than most spectroscopic catalogs.  

Because only 49 stars in the IRTF library have NIR photometry in the GCPD we supplement the IRTF spectra with NIR spectra from \citet{Mann2013a}, \citet{2013ApJ...779..188M}, and \citet{2014MNRAS.443.2561G}. All of these stars are K and M dwarfs, which biases the sample toward redder $B-V$ colors. However, the paucity of properly flux-calibrated NIR spectra makes this problem difficult to avoid. NIR spectra from these sources were taken with the same instrument and mode, and reduced in an identical manner to that of the IRTF library \citep{Vacca:2003qy, Cushing:2004fk}, except that the supplemental spectra cut off at 2.5\um\ while most stars in the IRTF library extend to 5\um. 

We remove stars known to be highly variable \citep[$\gtrsim0.1$~mag variability in $V$,][]{2004yCat.2250....0S, 2010OAP....23..102S}, stars lacking any photometry in the GCPD, those with an object of similar or greater brightness within 8\arcsec\ (which will contaminate the photometry), and stars with questionable designations (e.g., the star labeled as GJ 15B in the NGSL catalog is probably GJ 15A based on the spectrum and HST pointing). This leaves 364 stars from the NGSL catalog, 49 stars from the IRTF library, and 59 stars from our supplemental spectra.

\section{Photometric Systems}\label{sec:phsample}
We focus our efforts on photometry from the GCPD because it is the most common source of photometry for bright interferometry stars. Many less commonly-used systems in the GCPD, such as Alexander \citep{1981MNRAS.194..403J} and Oja \citep{Haggkvist1970}, lack photometry for stars in our spectroscopic sample. To determine the minimum number of spectra required we run our analysis (Section~\ref{sec:analysis}) on random subsets of the $UBV$ filter sample. We find that if the subset contains $\lesssim40$ spectra (independent of how many photometric points there are per spectrum) the resulting filter profile is frequently ($\gtrsim30\%$ of the time) outside the range of profiles generated by our analysis of the full set (see Section~\ref{sec:errors} for more details). We therefore require that to be included in our analysis a system/filter must have photometry for at least 40 stars in our spectroscopic sample and these $>=40$ stars span at least 1 magnitude in $B-V$ color. We exclude Johnson $R$ and $I$ from our tests. Johnson $I$ is not fully covered by either of our spectral libraries, and Cousins $R$ and $I$ (which we do test) are more commonly used compared to their Johnson counterparts. We list all photometric systems tested in this paper in Table~\ref{tab:phot}.

\begin{figure}[tbp] 
   \centering
   \includegraphics[width=0.48\textwidth]{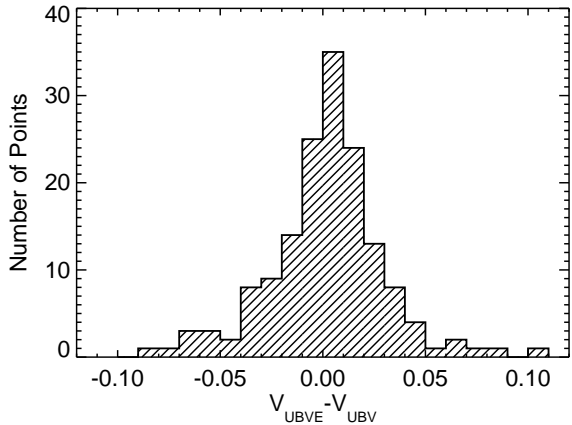} 
   \caption{Comparison of $V$ magnitudes from the GCPD $UBVE$ and $UBV$ systems. The median difference is 0.000, and the mean difference is 0.001 with a standard deviation of $\simeq$0.02, which is consistent with measurement noise and/or stellar variability. Much of the photometry in the GCPD listed under different names are actually on the same system. See Section~\ref{sec:phsample} for details.}
   \label{fig:comp_systems}
\end{figure}

The GCPD differentiates between photometric systems even though they are in fact indistinguishable. The Straizys system \citep{Kakaras:1968} and Vilnius systems are identical; $B$ and $V$ from the Cape (GCPD name $UcBV$) system are in the Johnson system; GCPD systems $UBV$ and $UBVE$ are indistinguishable (Figure~\ref{fig:comp_systems}); and (RI)C, Eggen, and $UBVRI$ all contain Johnson $U$, $B$, and/or $V$ measurements in addition to their own unique system measurements. For our purposes we analyze systems with identical or nearly identical parameters together. 

Initially we adopt the zero point and filter profile given in the GCPD (listed in Table~\ref{tab:phot}). Many of these systems have been subsequently revised \citep[e.g., ][]{2006AJ....131.1184M, 2011PASP..123.1442B, 2012PASP..124..140B}. However, our analysis only requires an approximate value for these as an initial guess. We compare our final filter profiles to updated literature values in Section~\ref{sec:results}.

\tabletypesize{\footnotesize}
\begin{deluxetable*}{l l l l l l l l l l l l l l}
\tablecaption{Photometric Systems and Initial Parameters}
\tablewidth{0pt}
\tablehead{
\colhead{GCPD Name} & \colhead{Band} & \colhead{System Reference(s)} & \colhead{Center} & \colhead{\weff$\tablenotemark{a}$}  & \colhead{N$_{\rm{sp}}$\tablenotemark{b}} & \colhead{N$_{\rm{ph}}$\tablenotemark{b}} \\
\colhead{} & \colhead{} & \colhead{} & \colhead{\AA} & \colhead{\AA}  & \colhead{} & \colhead{}
}
\startdata
$UcBV$ & $Uc$ & \citet{1958AJ.....63..118A} & 3948 &  376 & 56 &  71 \\
$UBV$ & $U$ & \citet{1951ApJ...114..522J}, & 3517 &  665 & 299 &  1708 \\
$UBVE$ & $B$ & \citet{1990PASP..102.1181B} & 4454 & 1037 & 314 & 2168 \\
 & $V$ &  & 5523 & 909 & 315 & 2373 \\
 %\hline
%$UBVRI$  & $R$ & \citet{1963BOTT....3..137M} & 7018 & 2052 & 126 &  209 \\
\hline
 & $u$ & & 3466 &  339 & 273 & 817 \\
$uvby$ & $b$ & \citet{1956VA......2.1336S}, & 4747 &  201 & 266 & 837 \\
(Str\"omgren) & $v$ & \citet{1966AJ.....71..114C} & 4073 &  200 & 266 & 818 \\
 & $y$ &  & 5479 &  256 & 267 & 882 \\
 \hline
 & $U$ & & 3467 &  479 & 269 & 267 \\
 & $B1$ &  & 4030 &  424 & 269 & 267 \\
 & $B$ & & 4266 &  806 & 269 & 267 \\
Geneva  & $B2$ & \citet{1972VA.....14...13G} & 4486 &  445 & 269 & 267 \\
 & $V1$ & & 5417 &  508 & 269 & 267 \\
  & $V$ & & 5524 &  779 & 269 & 267 \\
 & $G$ & & 5821 &  485 & 269 & 267 \\
 \hline
  & $U$ & & 3450 &  395 & 180 & 387 \\
 & $P$ & & 3740 &  276 & 180 & 387 \\
 & $X$ & & 4054 &  254 & 181 & 392 \\
Vilnius & $Y$ & \citet{Kakaras:1968} & 4665 &  288 & 180 & 392 \\
  & $Z$ & & 5162 &  232 & 180 & 392 \\
 & $V$ & & 5442 &  294 & 180 & 392 \\
 & $S$ & & 6534 &  212 & 179 & 377 \\
 \hline
   & $W$ & & 3546 &  453 & 171 & 171 \\
$WBVR$ & $B$ & \citet{1991TrSht..63....1K} & 4400 &  881 & 171 & 171 \\
 & $V$ & & 5531 &  924 & 171 & 171 \\
 & $R$ & & 7164 & 1312 & 171 & 171 \\
 \hline
    & $m35$ & & 3465 &  374 & 102 & 145 \\
 & $m38$ & & 3825 &  327 & 122 & 185 \\
DDO & $m41$ & \citet{1976AJ.....81..182M} & 4164 &   80 & 124 & 191 \\
 & $m42$ & & 4255 &   71 & 124 & 191 \\
 & $m45$ & & 4520 &   73 & 124 & 191 \\
 & $m48$ & & 4884 &  194 & 124 & 191 \\
 \hline
(RI)C & $R$ & \citet{1976MmRAS..81...25C} & 6507 & 1479 & 65 & 81 \\
 & $I$ & & 7904 & 1042 & 67 & 84 \\
 \hline
(RI)Eggen & $R$ & \citet{1965AJ.....70...19E} & 6350 & 1200 & 113 & 126 \\
 & $I$ & & 7900 & 900 & 113 & 94 \\
 \hline
& $J$ & \citet{1963BOTT....3..137M}, & 12602 & 3268 & 87  & 154 \\
$IJHKLMN$ & $H$ & \citet{Bessell:1988qy} & 16551 & 2607 & 89 & 144 \\
 & $K$ &  &  22094 & 5569  & 44 &  79\\
 \hline
 \hline
 Tycho$\tablenotemark{c}$ & $B_T$ & \citet{2012PASP..124..140B} & 4213 &  693 & 342 & 342\\
 & $V_T$ &  & 5345 & 1043 & 342 & 342\\
 \hline
\hipp$\tablenotemark{c}$	& $H_p$	 & \citet{2012PASP..124..140B} & 5593 & 2471 & 346 & 346
 \enddata
 \tablecomments{Details of the photometric sample can be found in Section~\ref{sec:phsample}.}
\tablenotetext{a}{The effective width (\weff) is defined as: \weff$ =  \left(\bigintss{S_x(\lambda) d\lambda}\right)/\rm{max}(S_x(\lambda)$). This is essentially the equivalent width divided by the maximum throughput within the filter bandpass.} 
\tablenotetext{b}{N$_{\rm{ph}}$ is the number of photometric points included in our analysis, N$_{\rm{st}}$ is the number of spectra with measurements in that filter. N$_{\rm{ph}}\ge$\,N$_{\rm{sp}}$ because many stars have more than one measurement in a given band.}
\tablenotetext{c}{Tycho, and \hipp\ not in the GCPD, so we list their common name.}
\label{tab:phot}
\end{deluxetable*}

In addition to the photometry listed in Table~\ref{tab:phot} we  analyze photometry from \hipp{} \citep{van-Leeuwen:2005kx, van-Leeuwen:2007yq}, and Tycho-2 \citep{Hog2000}. We download 2MASS photometry (for the NIR data), although we do not attempt to derive new zero points or filter profiles for 2MASS, as the calibration of our NIR data is not sufficiently precise to improve upon the calibration from \citet{2003AJ....126.1090C}. Rather, 2MASS data is only included to improve the absolute flux calibration of the NIR spectra. For \hipp{} and Tycho-2 photometry we use the initial zero points and system sensitivities from \citet{2012PASP..124..140B} and for 2MASS we use those from \citet{1992AJ....104.1650C,2003AJ....126.1090C}.

As noted by \citet{2012PASP..124..140B} and \citet{Casagrande2014}, there is some confusion in the literature regarding photonic (photon-counting) versus energy response functions. Since the bands are generally normalized the difference is a factor of $\bigintss{\lambda d\lambda}$ (see Equations~\ref{eqn:synthflux} and \ref{eqn:QE} below), which can be very significant for broadband photometry. For references that provide ambiguous information we assume they report photonic filter profiles, as we found this gave better agreement with our final results. We convert all photonic response curves to energy response functions (more appropriately called system sensitivity, which we denote as $S_x$). The energy response is easier to use when calculating synthetic fluxes from flux-calibrated spectra (generally in energy units). For our derived filter profiles (Section~\ref{sec:results}, Appendix A) we report energy response functions.

Photometry for each star is downloaded from the GCPD using the python script \textit{GCPD3}\footnote{\href{https://github.com/awmann/GCPD3}{https://github.com/awmann/GCPD3}}. Photometry from \hipp, Tycho-2, and 2MASS (with errors) are downloaded using the IDL routine \textit{QUERYVIZIER}\footnote{\href{http://idlastro.gsfc.nasa.gov/ftp/pro/sockets/queryvizier.pro}{http://idlastro.gsfc.nasa.gov/ftp/pro/sockets/queryvizier.pro}}.

\section{Analysis}\label{sec:analysis}
Here we compare the literature photometry with the flux calibrated spectra and derive updated filter profiles, zero points, and photometric corrections.

\subsection{Comparing photometric fluxes to spectroscopic fluxes}\label{sec:compare}
The synthetic flux ($f_{syn,x}$) is the flux in the sample spectrum, integrated over the filter profile, where the transmission of the filter is a wavelength dependent quantity, and is calculated using the formula:
\begin{equation}\label{eqn:synthflux}
f_{syn,x}  =  \frac{\bigintss{f_{\lambda}(\lambda) S_x(\lambda) d\lambda}} {\bigintss{S_x(\lambda) d\lambda}}, 
\end{equation}
where $f_{\lambda}$ is the flux-calibrated spectrum\footnote{The more appropriate term for $f$ is radiative flux density or spectral irradiance, but for simplicity we use the more common astronomy term, flux or spectrum.} (e.g., $\rm{erg}\,\rm{cm}^{-2}\,s^{-1}\angstrom^{-1}$) as a function of wavelength ($\lambda$). $S_x(\lambda)$ is the system's {\it energy} sensitivity for the filter $x$ (e.g., $U$, $B$, $V$), defined as:
\begin{equation}\label{eqn:QE}
S(\lambda) = QE(\lambda)\lambda,
\end{equation}
where $QE(\lambda)$ is the unit-less fractional transmission (i.e., photons detected/photons received). The combination of Equations~\ref{eqn:synthflux} and \ref{eqn:QE} put us in accord with Equation~A11 from \citet{2012PASP..124..140B} and Equation~3 from \citet{Bohlin2014}.
All synthetic fluxes are calculated using custom IDL routines, but we test to ensure consistency with the commonly used python tool \textit{Pysynphot}. 

Corresponding photometric fluxes from GCPD, \hipp, Tycho-2, and 2MASS photometry (denoted by $f_{phot,x}$) are computed using the formula:
\begin{equation}
f_{phot,x} = f_{0,x}10^{-0.4(m_x)} 
\end{equation}
where $m_x$ is the magnitude and $f_{0,x}$ is the zero point for band $x$. For this work we use flux units (erg\,cm$^{-1}\,s^{-2}$\,\AA$^{-1}$) rather than magnitudes when discussing zero points because they are more intuitive when discussing fractional differences (e.g., 30\% is easier to imagine than 0.39 magnitudes) and because they are more practical for absolutely flux calibrating spectra. We report magnitudes where possible for reference.

\begin{figure*}[tbp] 
   \centering
   \includegraphics[width=0.47\textwidth]{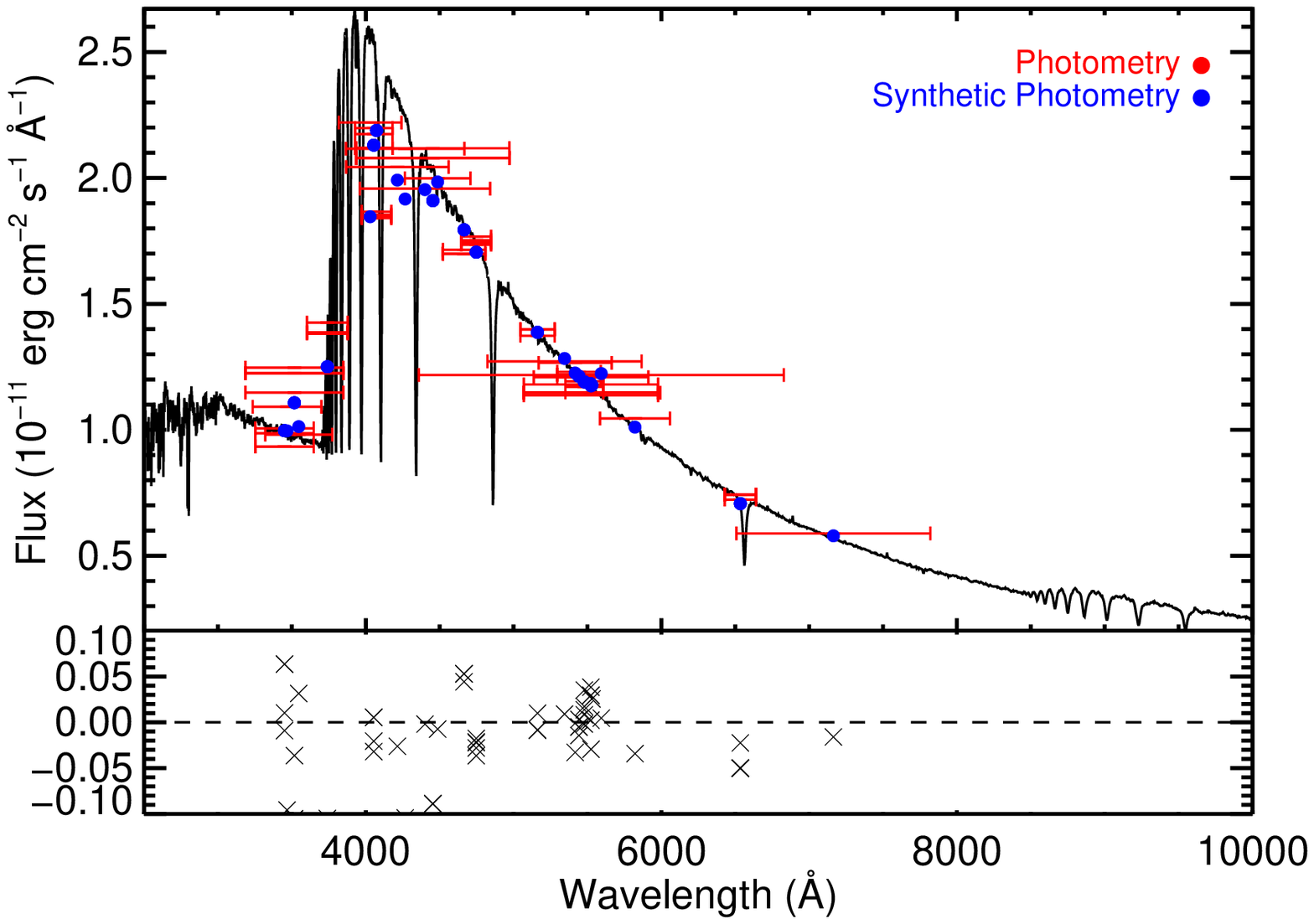} 
   \includegraphics[width=0.47\textwidth]{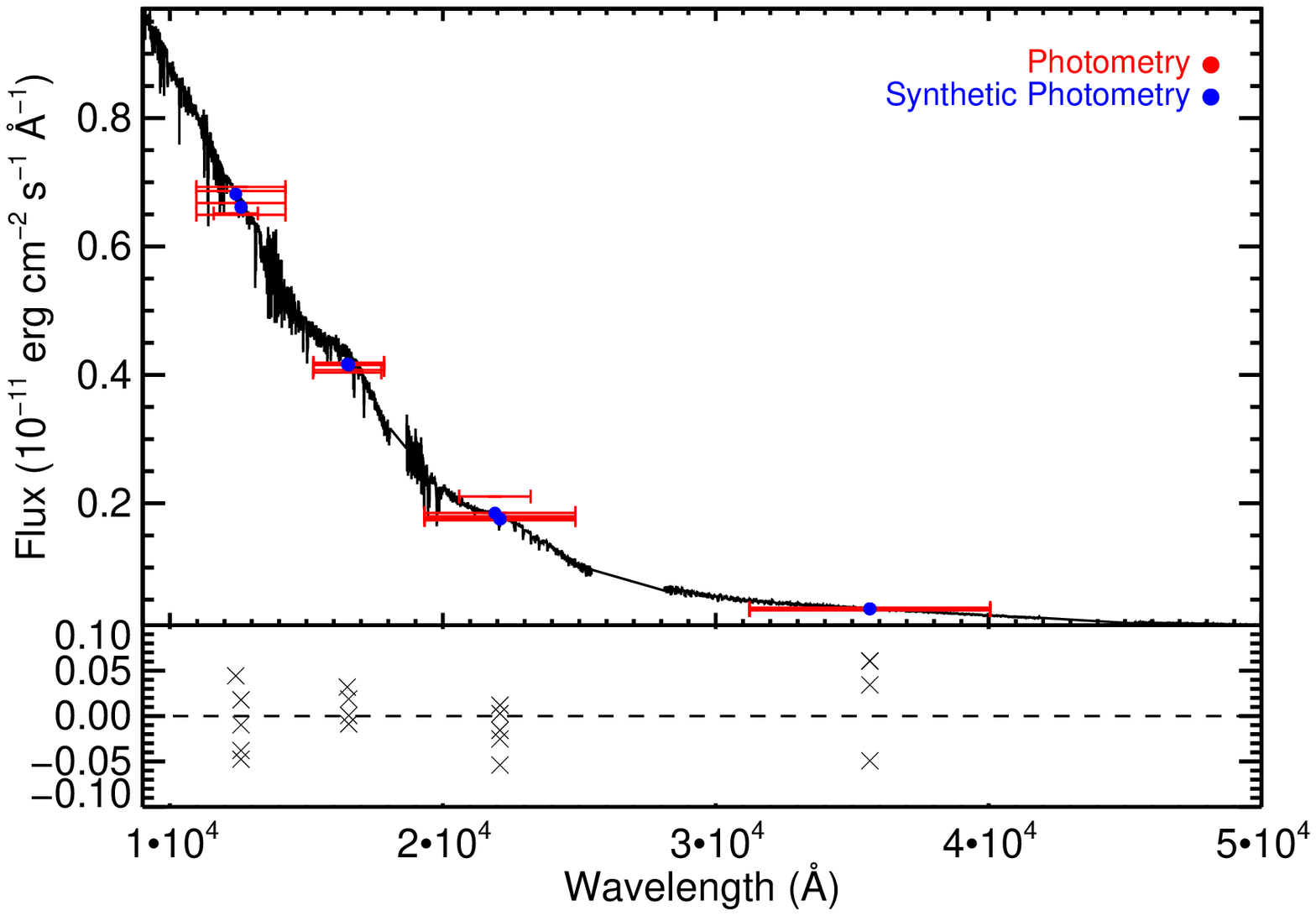} 
   \caption{Demonstration of our flux calibration of the NGSL spectrum of the B9 dwarf HD 147550 (left) and the IRTF library spectrum of the M2 dwarf HD 95735 (right). Flux derived from photometry is shown in red with `error bars' showing the approximate FWHM of the filter profile. Blue circles indicate the corresponding flux estimated from the spectrum (synthetic flux). Fractional residuals (synthetic - photometric)/synthetic are shown on the bottom panel of each figure. See Section~\ref{sec:compare} for more details.}
   \label{fig:fluxcal}
\end{figure*}

In most cases, spectrophotometric libraries have precise {\it relative} flux calibrations (i.e., flux levels as a function of wavelength) but can still have significant absolute flux calibration errors (i.e., the whole spectrum is off by a constant). \citet{2012PASP..124..140B} solve this by renormalizing their spectra using precise \hipp\ ($H_P$) magnitudes. Instead, we renormalize using all available photometry, which is also more resistant against outlier photometry (e.g., taken during a flare) and enables us to include stars lacking $H_P$ magnitudes. Specifically, after we compute synthetic and photometric fluxes we multiply each spectrum by a constant, $C$, which we define as:
\begin{equation}\label{eqn:norm}
C = \left<f_{phot,x}/f_{syn,x}\right>,
\end{equation}
where $\left< \right>$ denotes the median, which we calculate over all bands ($x$) for a given star. This serves to absolutely flux calibrate each spectrum.  The 1$\sigma$ error on $C$ is assumed to be the standard deviation of $f_{phot,x}/f_{syn,x}$ values after removing 5$\sigma$ outliers. Note that although we are calibrating the spectrum using uncorrected (and possibly erroneous) photometry, our process for fixing the photometric parameters is iterative and converges quickly (see Section~\ref{sec:new} for more details). We show two example calibrated spectra in Figure~\ref{fig:fluxcal} with the photometric fluxes for comparison. 

In theory it is possible to get a better $C$ (and error estimate) using a robust weighted mean instead of a median. This is impractical because: a) measurement and zero point errors (and hence weights) for the GCPD photometry are not known, b) while Tycho-2 and \hipp{} report measurement errors, zero point errors are not known, and 3) a median does a better job at ignoring outlier photometry when there are only $\simeq$5 measurements. 

\begin{figure}[tbp] 
   \centering
   \includegraphics[width=0.5\textwidth]{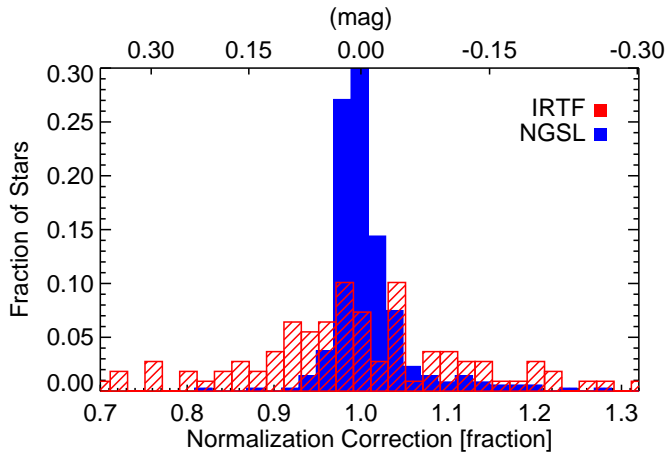} 
   \caption{Distribution of normalizations applied to the NGSL and the IRTF library with the supplemental SpeX spectra. Five spectra have corrections $>30\%$ land outside the range of this plot, three of which are from our supplemental IRTF data. Changes in flux (as a fraction) are shown on the bottom X-axis with corrections in magnitudes on top for reference. See Section~\ref{sec:compare} for more details on our procedure to absolutely flux calibrate the spectra.}
   \label{fig:norms}
\end{figure}

We show the distribution of corrections ($C$) applied to each of our libraries in Figure~\ref{fig:norms}. The distribution similar to that derived for the NGSL library by \citet{2012PASP..124..140B}, who renormalize their spectra using \hipp{} photometry alone. In general the corrections are small for the NGSL library, but there are a number of outliers with $>20\%$ corrections. Most of these outliers are also seen in the distribution of renormalizations from \citet{2012PASP..124..140B}, although many they identify appear to be cut by our variability criterion (See Section~\ref{sec:spsample}). Corrections for the IRTF library often exceed 10\%. Estimated errors on the normalization are also generally larger for the IRTF data ($\sim$1--3\%) than for the NGSL library ($<1\%$). This is because there are typically only 5--6 photometric points for a given star in the NIR (2MASS $JHK_{\rm{S}}$ and Johnson $JHK$) while most of the NGSL stars have $>30$ photometric measurements, and because of lower precision flux calibration for the IRTF library.

Once the spectrum has been renormalized, we recalculate all synthetic fluxes following Equation~\ref{eqn:synthflux}. We then examine the ratio of photometric to synthetic flux ($f_{phot,x}/f_{syn,x}$) as a function of $B-V$. $B-V$ magnitudes for each star are computed Tycho-2 $B_T$ and $V_T$ using the formula from \citet{2002AJ....124.1670M}, where available. For stars lacking Tycho-2 magnitudes we use $B-V$ colors from the GCPD. We use $B-V$ as a measure of color because it is the only color index that is available for all stars in our spectroscopic sample. 

\begin{figure*}[tbp] 
   \centering
   \includegraphics[width=0.475\textwidth]{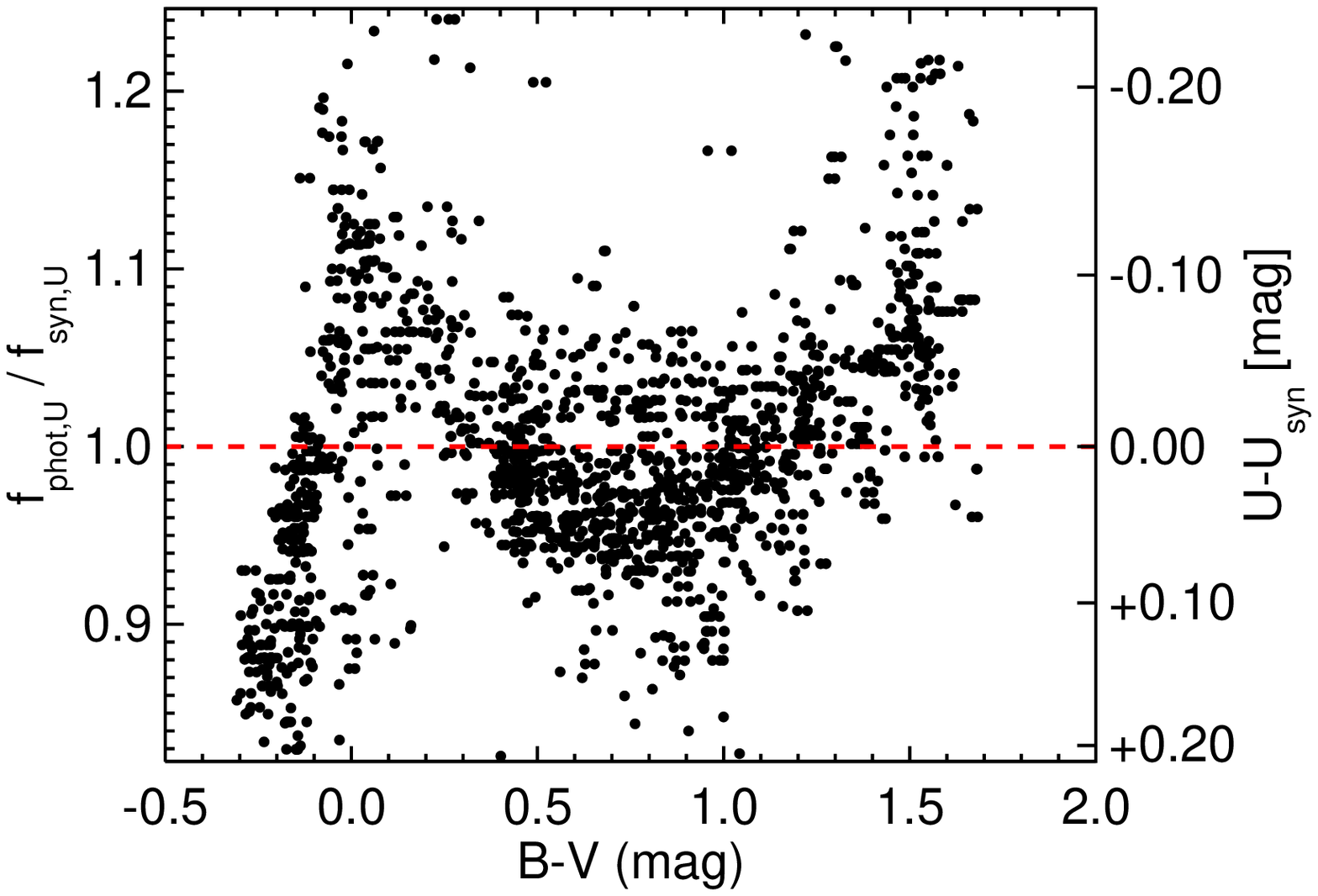} 
   \includegraphics[width=0.475\textwidth]{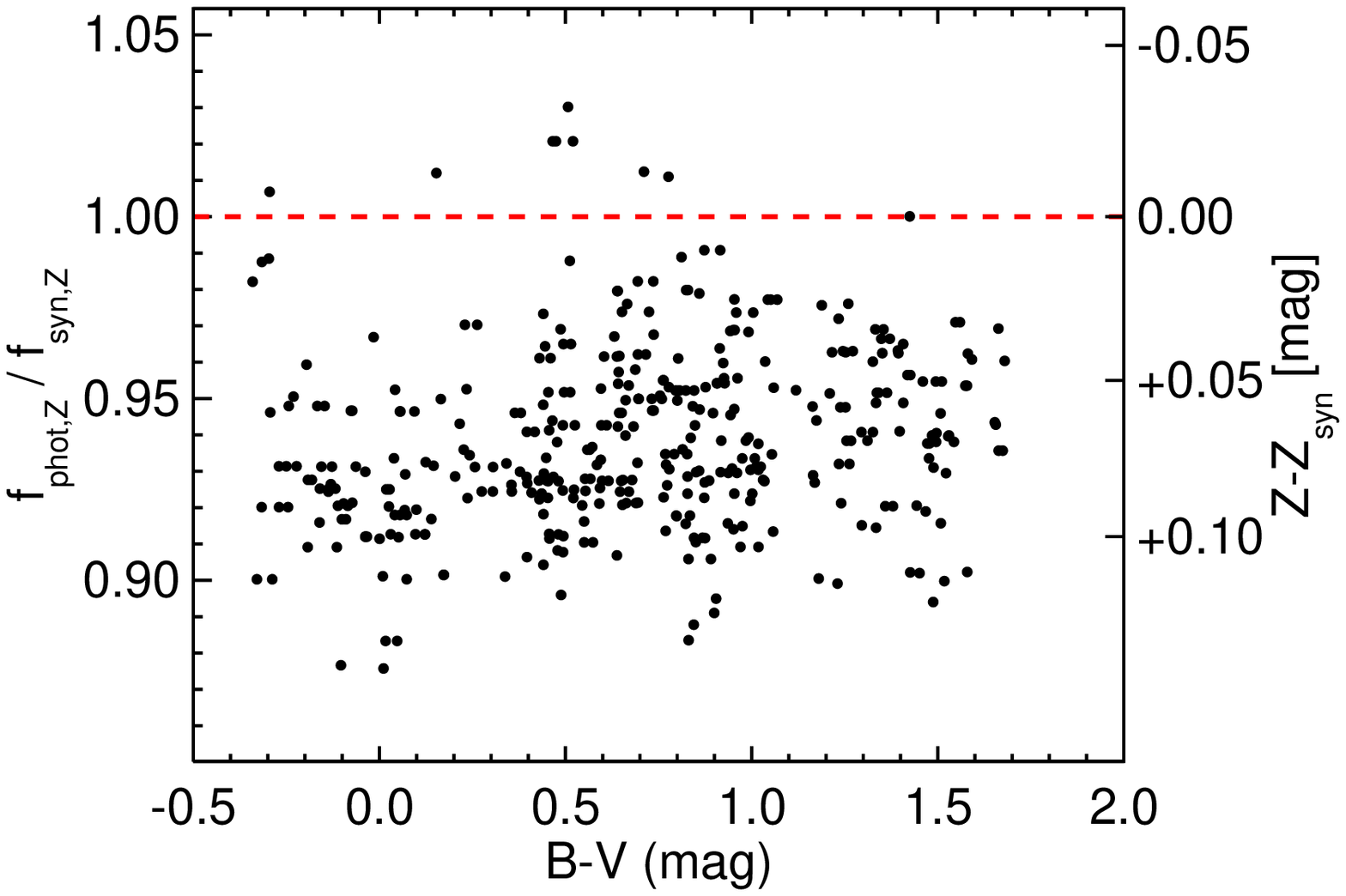} 
   \caption{The ratio of flux from observed photometry ($f_{phot}$) to that calculated from the spectra ($f_{syn}$) as a function of $B-V$ color for Johnson $U$ (left) and Vilnius $Z$ (right). $B-V$ magnitudes are calculated from Tycho-2 $B_T$ and $V_T$ where available, and from the GCPD elsewhere. The magnitude differences is shown on the right Y-axis for reference. Note that the Y-axis scales of the two plots are not the same. The red dashed line indicates agreement between photometric and spectroscopic fluxes. We have added an artificial scatter of 0.025~magnitudes in $B-V$ to spread out overlapping points. See Section~\ref{sec:compare} for more information.}
   \label{fig:ratio1}
\end{figure*}

We find significant discrepancies between synthetic and photometric flux as a function of $B-V$ color (see Figure~\ref{fig:ratio1} for two examples). The shape and amplitude of the trend with $B-V$ varies for each filter/system, but tends to be worse for filters with larger wavelength coverage (broader). Differences between photometric and spectroscopic fluxes of 10--20\% are typical for broad filters and for the reddest and bluest stars. Both broad and narrow filters typically show systematic offsets ($\simeq$5--10\%) from a (median) flux ratio of unity. 

\begin{figure}[tbp] 
   \centering
   \includegraphics[width=0.5\textwidth]{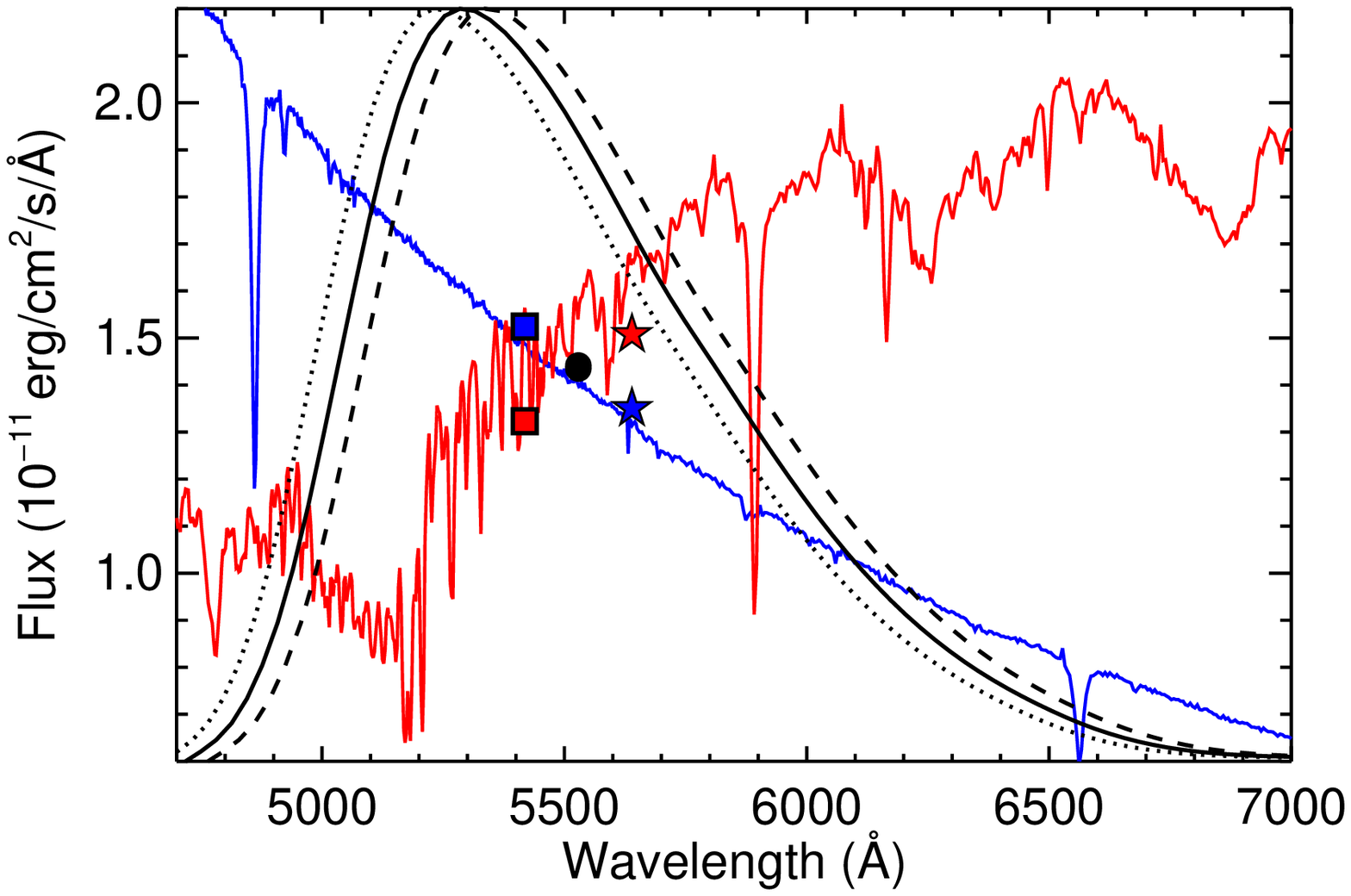} 
   \caption{Example of how slightly inaccurate filter profiles create color-terms. We show the spectrum of the B6 star HD 174959 (blue) and the K7 star HD 201092 (red). These stars are selected because they have nearly identical $V$ magnitudes. When we use the correct $V$ filter profile (solid black line) the derived fluxes for both stars are nearly identical (black dot). If we shift the filter passband 1\% redward (dashed line) the derived flux increases by $\simeq6\%$, but decreases by $\simeq6\%$ for the B6 star (red and blue stars). The effect has a similar magnitude but opposite sign when we shift the filter blueward. More details can be found in Section~\ref{sec:compare}}
   \label{fig:color}
\end{figure}

The trends we see in $f_{phot,x}/f_{syn,x}$ with $B-V$ can be explained by errors in the filter profile. For example, an optical filter profile that is redder than the true system throughput will include extra flux from an M star because the SED of the M star is increasing with wavelength over the filter passband, but miss flux from an A star because the SED is decreasing. We illustrate a simple version of this effect in Figure~\ref{fig:color}. More complicated trends like that seen in Figure~\ref{fig:ratio1}a are the consequence of multiple, overlapping issues with the filter profile. Errors in width, shape, and central wavelength of the filter passband combined with non-smooth spectra (e.g., a deep temperature sensitive feature somewhere in the passband) can give rise to such polynomial relations. We note that most filters show trends significantly closer to linear than what we see with Johnson $U$ shown in Figure~\ref{fig:ratio1}. 

Although the trends are large they are also relatively tight, specifically, the scatter around a polynomial fit to the data is small. An examination of the outlier points indicates that many of the stars have additional discrepant photometry in the same filter, suggesting stellar variability or erroneous measurements can explain much of the scatter. Although we screened for known variables and tighter binaries, many will be missed. For statistical purposes we do not remove any of these outlier points.

\begin{figure}[tbp] 
   \centering
   \includegraphics[width=0.5\textwidth]{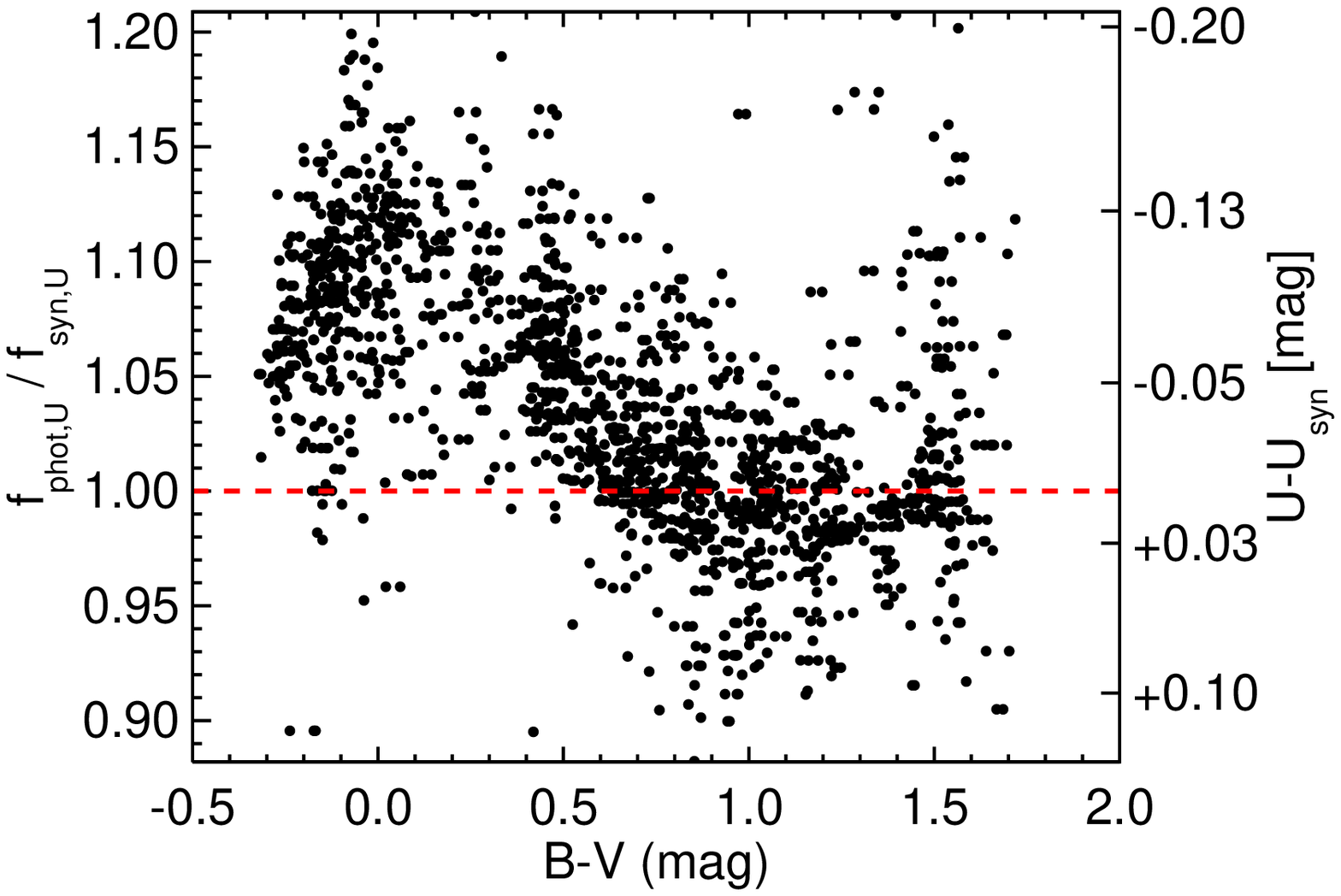} 
   \caption{Similar to Figure~\ref{fig:ratio1} for the Johnson $U$ band, but using the zero points and passbands are taken from \citet{2012PASP..124..140B}. This is an improvement over the older system parameters but a significant trend is still seen. Note that the Y-axis scale is similar but not identical to that in Figure~\ref{fig:ratio1}.}
   \label{fig:bessell_trend}
\end{figure}

The disagreement between photometric and spectroscopic fluxes decreases if we adopt more recent system parameters from the literature. For example, if we use filter profiles and zero points from \citet{2012PASP..124..140B} for the Johnson $U$, $B$, and $V$ filters the $B$ and $V$ synthetic and photometry-based fluxes agree at the level of $\simeq$2-6\%, compared to 10--15\% when using the GCPD parameters. Small trends with $B-V$ color are still present but only amount to 2--6\% changes over 1.5 magnitudes in $B-V$. The improvement in $U$ is much less significant (Figure~\ref{fig:bessell_trend}). This is most likely because \citet{2012PASP..124..140B} make use of the MILES library, which does not cover all of the $U$ band. Instead they have to interpolate past the end of their spectra, which introduces error.

\subsection{Deriving new system parameters}\label{sec:new}
\begin{figure}[tbp] 
   \centering
   \includegraphics[width=0.5\textwidth]{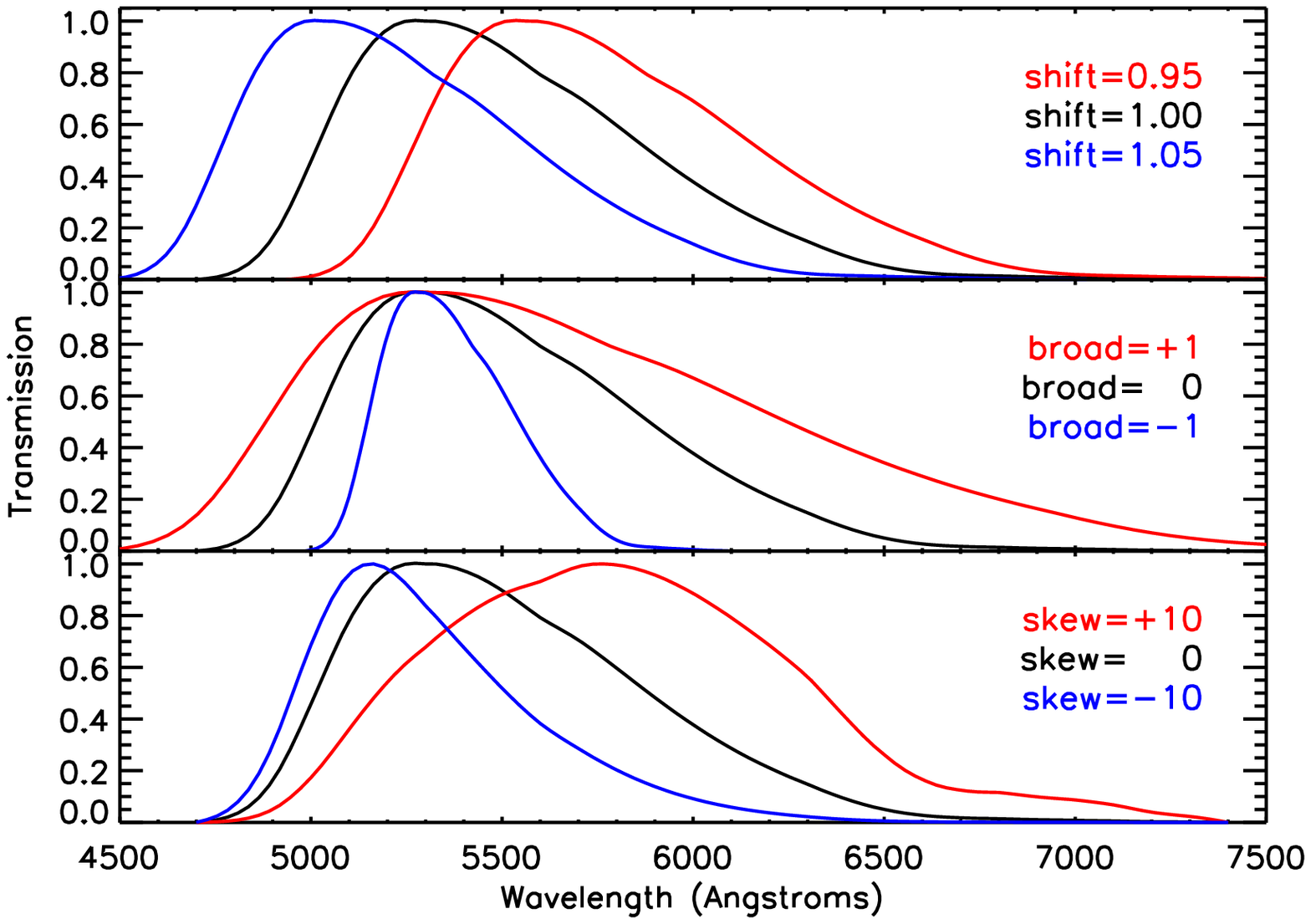} 
   \caption{The effects of Equations (5)-(7) on the filter profile. $Shift$ moves the central wavelength, $broad$ adjusts how narrow or broad the profile is, and $skew$ changes relative transmission between the blue and red end of the profile. Relatively large values of each parameter are used for this Figure to make the effect more clear. }
   \label{fig:filter_params}
\end{figure}

\begin{figure*}[tbp] 
   \centering
   \includegraphics[width=0.48\textwidth]{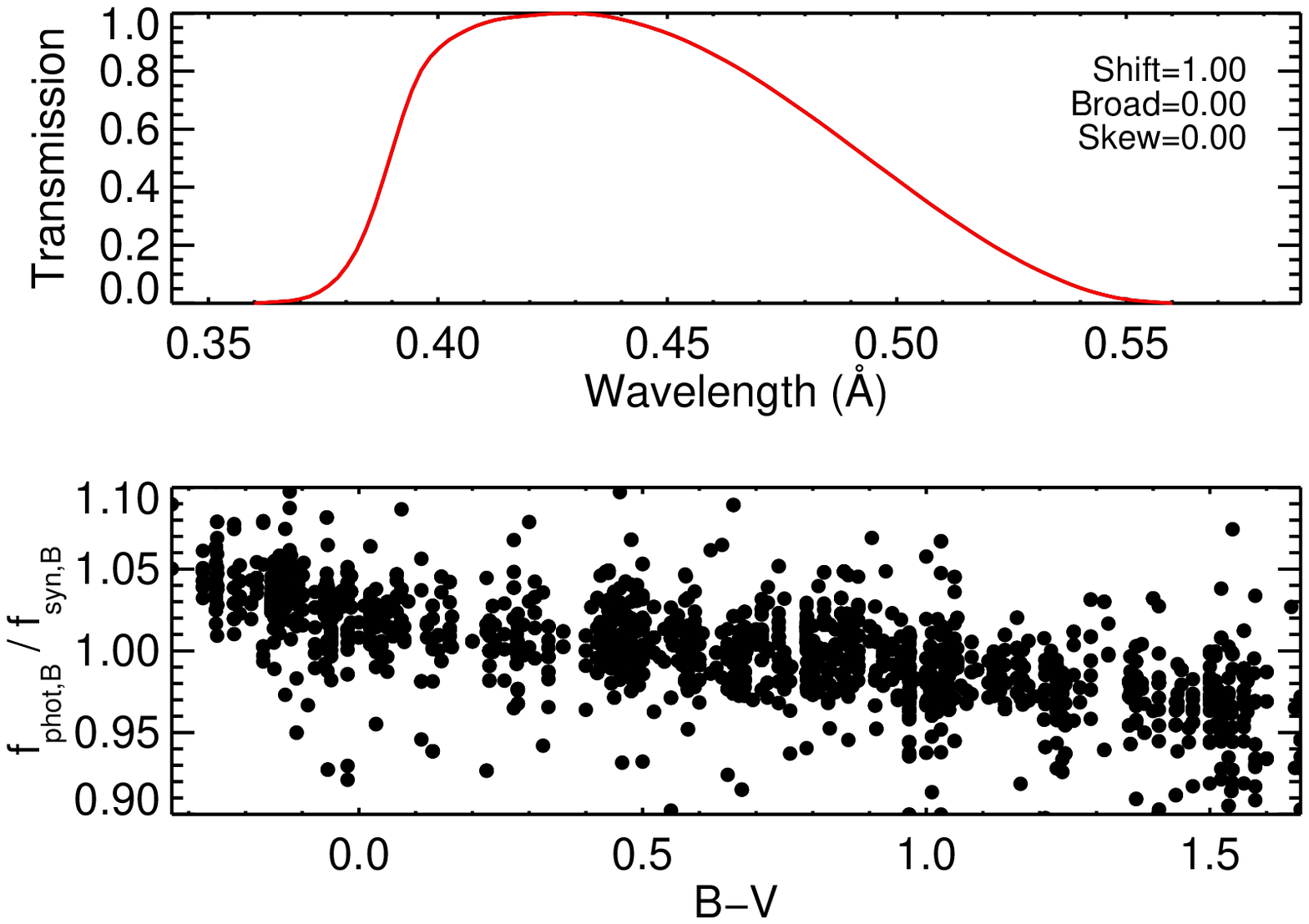} 
   \includegraphics[width=0.48\textwidth]{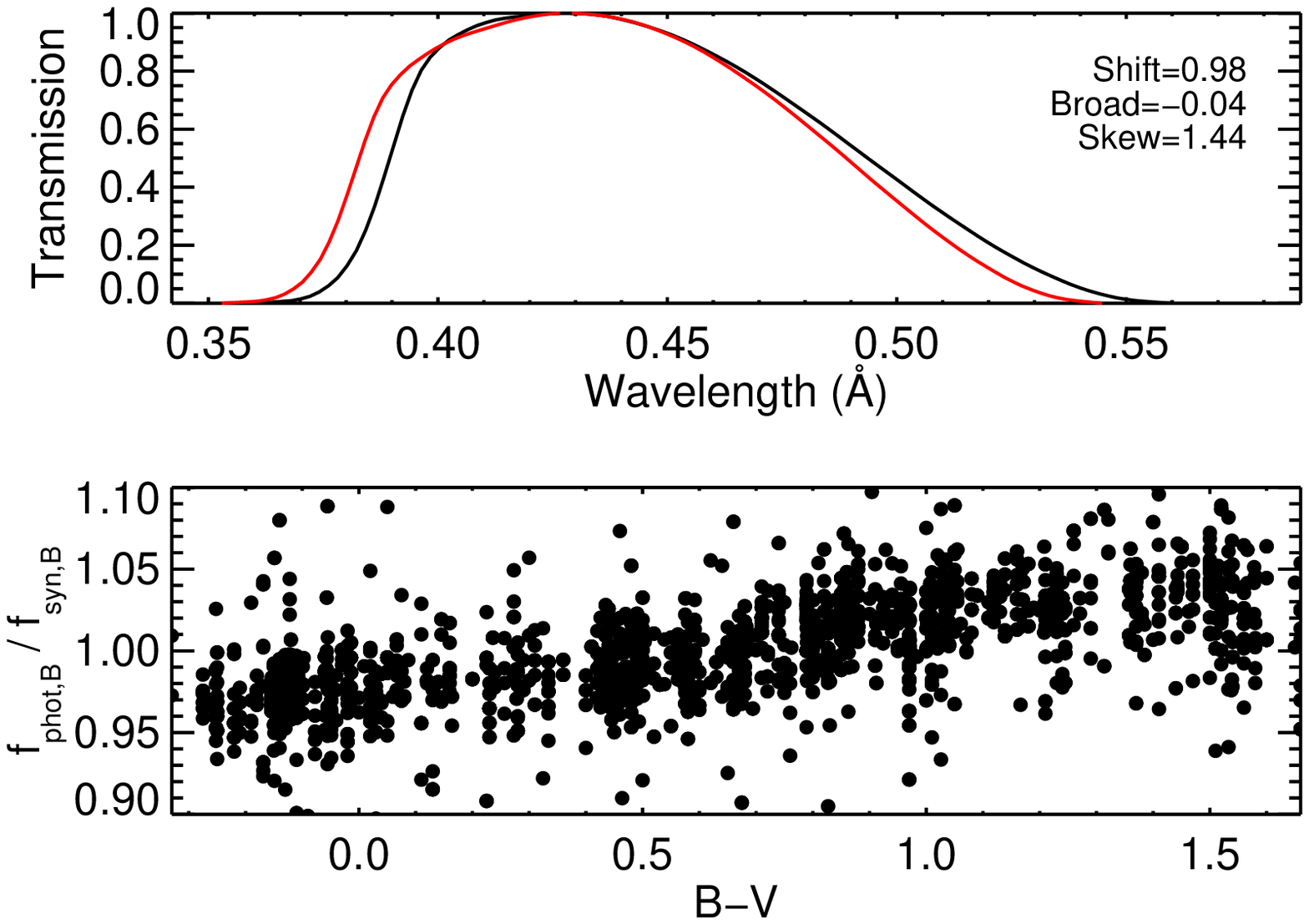} 
   \includegraphics[width=0.48\textwidth]{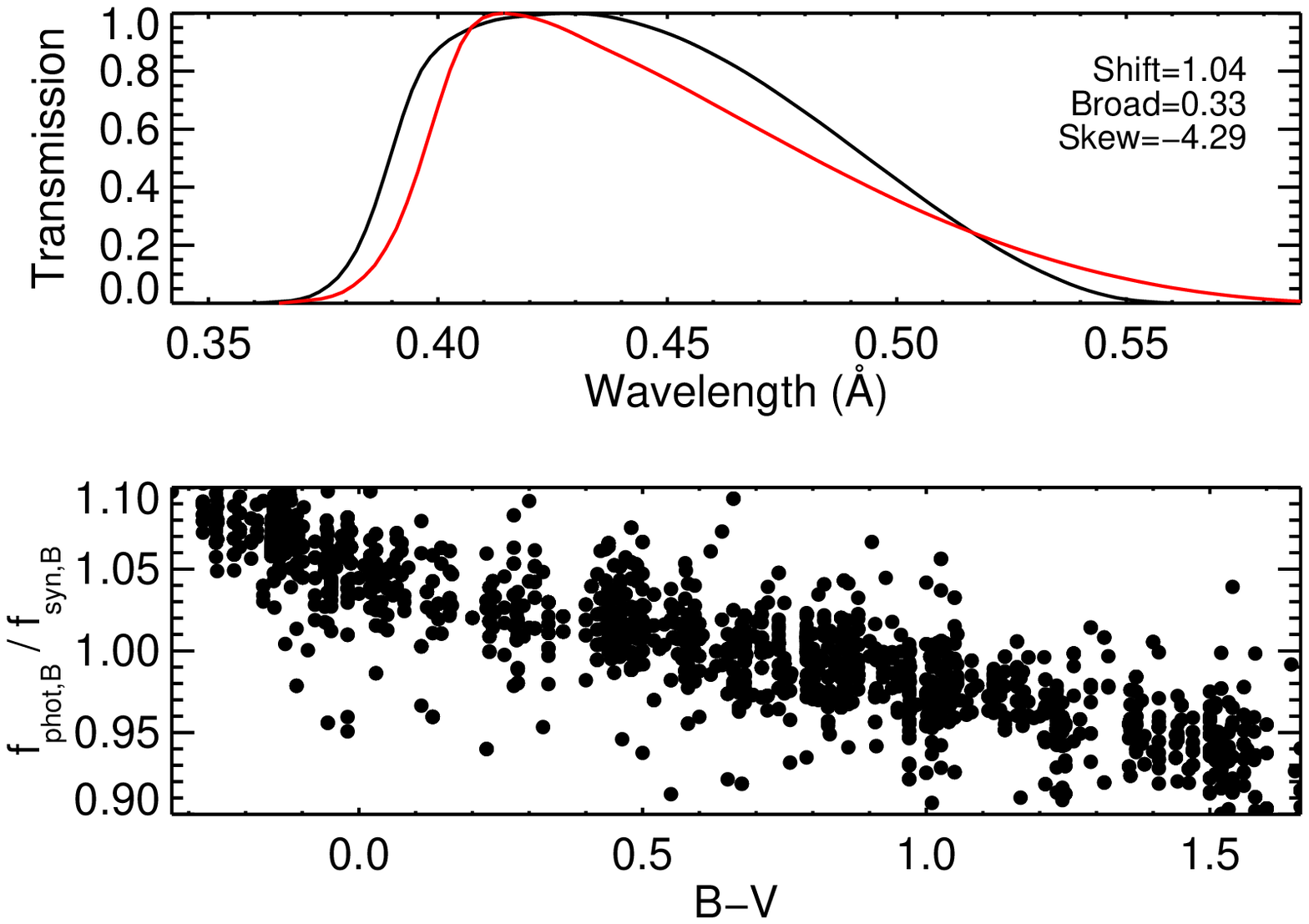} 
   \includegraphics[width=0.48\textwidth]{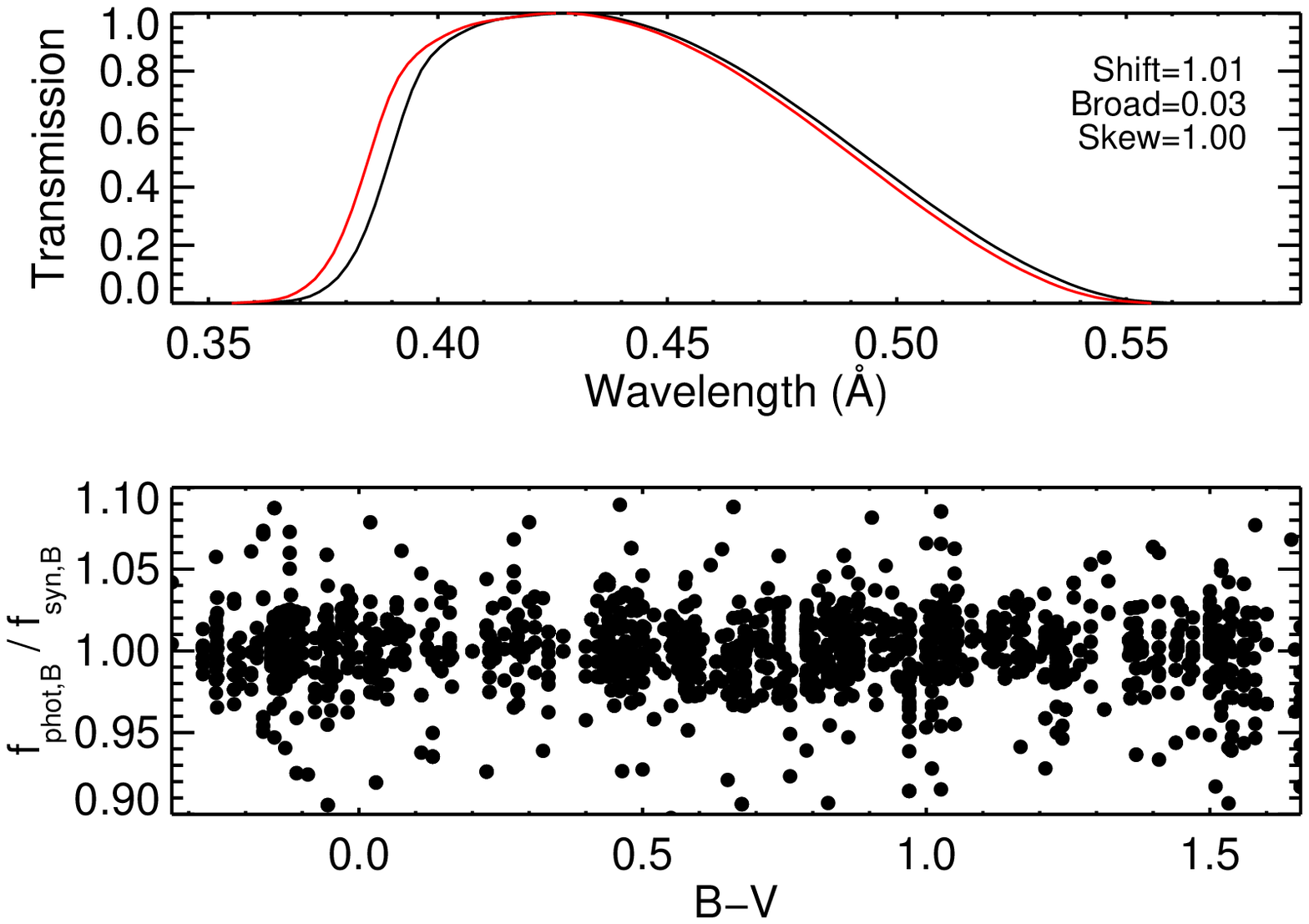} 
   \caption{Four example steps in the MC chain for the Johnson $B$ filter. The top of each pane shows the filter profile adopted by that step in red, with the initial input profile in black, along with the $skew$, $shift$, and $broad$ parameters in the legend. The bottom of each pane is analogous to Figures~\ref{fig:ratio1} and \ref{fig:bessell_trend}. Each pane represents a different step in the MC, with the top-left showing the initial value, and the bottom right showing the final result, and the other top right and bottom left displaying representative spots along the chain. Axis ranges are identical in all plots. Our MC procedure is described in Section~\ref{sec:new}.}
   \label{fig:MC}
\end{figure*}

We derive a new passbands and zero points for each band using a Monte Carlo (MC):
\begin{itemize}
\item Adjust the filter profile's $shift$, $skew$, and $broad$, which we define according to the following equations:
\begin{eqnarray}
 \lambda_{mod} &=& \lambda*shift       \\
 \lambda_{mod} &=& \lambda*((\lambda-\lambda_{peak})+1)^{broad} \\
 S_{mod}(\lambda) &=& S(\lambda)*\lambda^{skew},
\end{eqnarray}
where $\lambda_{mod}$ and $S_{mod}$ are the wavelength and flux levels for the modified filter profile, $\lambda_{peak}$ is the wavelength of maximum flux transmission. Figure~\ref{fig:filter_params} shows how each of these parameters changes the overall profile.
\item Normalize filter profiles such that their maximum transmission is unity.
\item Recalculate synthetic fluxes ($f_{syn,x}$) for all stars from the calibrated spectra and the modified filter profile.
\item Flag and remove points with $f_{phot,x}/f_{syn,x}$ more than 5$\sigma$ outside a 20-point running robust mean. 
\item Calculate and apply a correction to zero point such that the median $f_{phot,x}/f_{syn,x}$ is 1.
\item Calculate the probability that $B-V$ is correlated with $f_{phot,x}/f_{syn,x}$ using a Spearman rank test \citep{spearman04}.
\item Compute the RMS scatter in $f_{phot,x}/f_{syn,x}$. 
\end{itemize}
The above steps are repeated, each time adjusting the $shift$, $skew$, and/or $broad$ parameters, which in turn alter the filter profile. $Shift$ is varied from 0.9 to 1.1, $broad$ from $-2$ to 2, and $skew$ from $-5$ to 5, all in increments of 0.01. The range of values for $shift$, $skew$ and $broad$ are conservatively large (by design); none of our final filter profiles are near the limits for any parameter. After the MC is complete, we adopt the filter parameters that yield a $<50\%$ probability of a correlation with $B-V$ and the smallest scatter in $f_{phot,x}/f_{syn,x}$. We show illustrative examples of steps in the MC chain in Figure~\ref{fig:MC}.

The Spearman rank test is best at identifying linear trends in the data. If there are significant non-linear trends in the data, a Spearman rank test might find a low probability of correlation even for highly correlated data. We remedy this by splitting the data into two subsets along the median $B-V$ color and then recalculate the Spearman rank coefficient on each subset and for each MC step. We throw out profiles that show statistically significant correlation in either subset. Examination of remaining $f_{phot,x}/f_{syn,x}$ versus $B-V$ color plots suggests this method is sufficient to remove obvious poor profiles.

The absolute flux calibration factors we apply to each spectrum (see Section~\ref{sec:compare}, Figure~\ref{fig:norms}) are based on uncorrected photometry. Although $C$ (Equation~\ref{eqn:norm}) is calculated from the median of many points, the normalizations will still be slightly off due to errors in the filter parameters. To remedy this, we re-derive the absolute flux calibration constant for each library spectrum using the updated filter parameters. The changes are generally small, with a median normalization change of 0.3\%, but the largest changes are $>$10\% (stars with few measurements or most measurements in the same filter), and show a trend with stellar color (the reddest and bluest stars change the most). After renormalizing the spectra with updated filter parameters we rerun the MC, deriving new filter profiles and zero points, this time with better-calibrated spectra. This process is repeated until changes in the derived zero points and filter profiles are negligible ($<0.1\%$). For most filters this happens after just 2--3 iterations, with the worst cases taking 5 iterations.

\section{Results}\label{sec:results}
\subsection{Updated Passbands and zero points}\label{sec:newcal}
We list our final derived parameters for each system in Table~\ref{tab:final}. Zero points are given in erg\,cm$^{-1}\,s^{-2}$\,\AA$^{-1}$, but can be converted to magnitudes using the formula:
\begin{equation}
ZP_{mag} = -2.5\log(ZP_{flux})-21.10.
\end{equation}
Full normalized filter curves are given in the Appendix. 

\begin{deluxetable*}{l l | l l r r}
\tablecaption{Final zero points and Passband Parameters}
\tablewidth{0pt}
\tablehead{
\colhead{Name} & \colhead{Band} & \colhead{zero point} & \colhead{$\pm$} & \colhead{Center} & \colhead{\weff}   \\
\colhead{} & \colhead{} & \multicolumn{2}{c}{$10^{-9}$\,erg\,cm$^{-1}\,s^{-2}$\,\AA$^{-1}$}  & \multicolumn{2}{c}{\AA} }
 \startdata
Cape & U$_{\rm{C}}$ &      23.25 &       0.14 &  3925 &   999\\
\hline
 & U &      4.264 &      0.022 &  3620 &  1380\\
 & B &      6.459 &      0.032 &  4412 &  1816\\
Johnson & V &      3.735 &      0.019 &  5529 &  1129\\
% & R &      1.691 &      0.009 &  7112 &  1644\\
 & J &      0.310 &      0.003 & 12603 &  2095\\
 & H &      0.113 &      0.001 & 16552 &  1362\\
 & K &     0.0400 &     0.0005 & 22094 &  2142\\
 \hline
 & u &      11.72 &       0.06 &  3485 &   950\\
Stromgren & b &      5.974 &      0.030 &  4671 &   338\\
 & v &      8.691 &      0.044 &  4125 &   513\\
 & y &      3.778 &      0.019 &  5474 &   596\\
\hline
 & U &      5.612 &      0.029 &  3453 &  1404\\
 & B1 &      6.676 &      0.034 &  4031 &   861\\
 & B &      2.838 &      0.014 &  4230 &  1523\\
Geneva & B2 &      10.76 &       0.05 &  4448 &   597\\
 & V1 &      7.460 &      0.038 &  5379 &   788\\
 & V &      3.728 &      0.019 &  5495 &  1220\\
 & G &      9.462 &      0.048 &  5785 &   543\\
\hline
 & U &      18.75 &       0.10 &  3478 &   884\\
 & P &      14.30 &       0.08 &  3750 &   418\\
 & X &      11.28 &       0.06 &  4061 &   584\\
Vilnius & Y &      6.792 &      0.035 &  4697 &   725\\
 & Z &      4.583 &      0.023 &  5204 &   244\\
 & V &      3.772 &      0.019 &  5464 &   584\\
 & S &      1.863 &      0.010 &  6520 &   271\\
\hline
 & W &      3.523 &      0.018 &  3554 &   963\\
WBVR & B &      6.619 &      0.034 &  4382 &  1495\\
 & V &      3.730 &      0.019 &  5519 &  1394\\
 & R &      1.698 &      0.009 &  7166 &  1360\\
\hline
 & m35 &      13.02 &       0.07 &  3521 &  1015\\
 & m38 &      6.867 &      0.037 &  3864 &   709\\
DDO & m41 &      21.16 &       0.11 &  4182 &   211\\
 & m42 &      19.97 &       0.10 &  4257 &   145\\
 & m45 &      13.15 &       0.07 &  4512 &   141\\
 & m48 &      4.966 &      0.025 &  4912 &   437\\
\hline
Cousins & R &      2.215 &      0.012 &  6615 &  1877\\
 & I &      1.163 &      0.007 &  8047 &  1604\\
\hline
Eggen & R &      2.293 &      0.016 &  6989 &  2831\\
 & I &      1.337 &      0.011 &  8492 &  1330\\
\hline
Tycho & Bt &      6.798 &      0.034 &  4220 &  1455\\
 & Vt &      4.029 &      0.020 &  5350 &  1665\\
\hline
Hipparcos & Hp &      3.926 &      0.020 &  5586 &  2569
\enddata
\tablecomments{Zero points are all derived from our MC analysis (see Section~\ref{sec:analysis}). Errors are calculated from estimates of systematic errors in the flux calibration of our spectra and the standard error (see Section~\ref{sec:errors}). Full filter profiles are given in the Appendix. }
\label{tab:final}
\end{deluxetable*}

For Johnson $JHK$ filters our derived passbands are indistinguishable from the original. Compared to other systems we examine, there is a paucity of measurements for $JHK$ that overlap with our spectroscopic sample. Given this, and the larger errors associated with flux calibration of NIR data we decide to adopt the original filter profiles, but still adjust the zero points for these three filters.

\begin{figure*}[tbp] 
   \centering
   \includegraphics[width=0.475\textwidth]{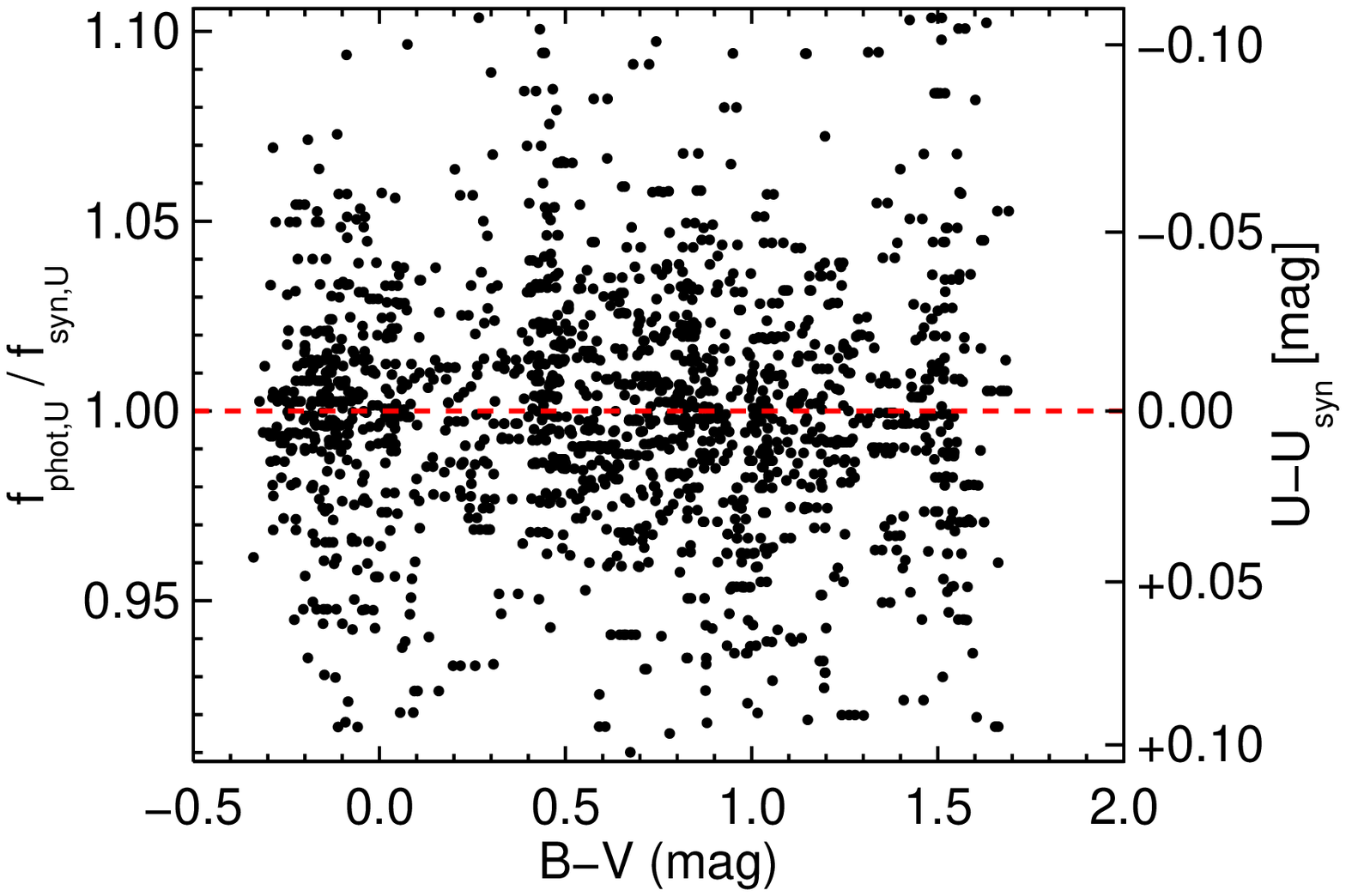} 
   \includegraphics[width=0.475\textwidth]{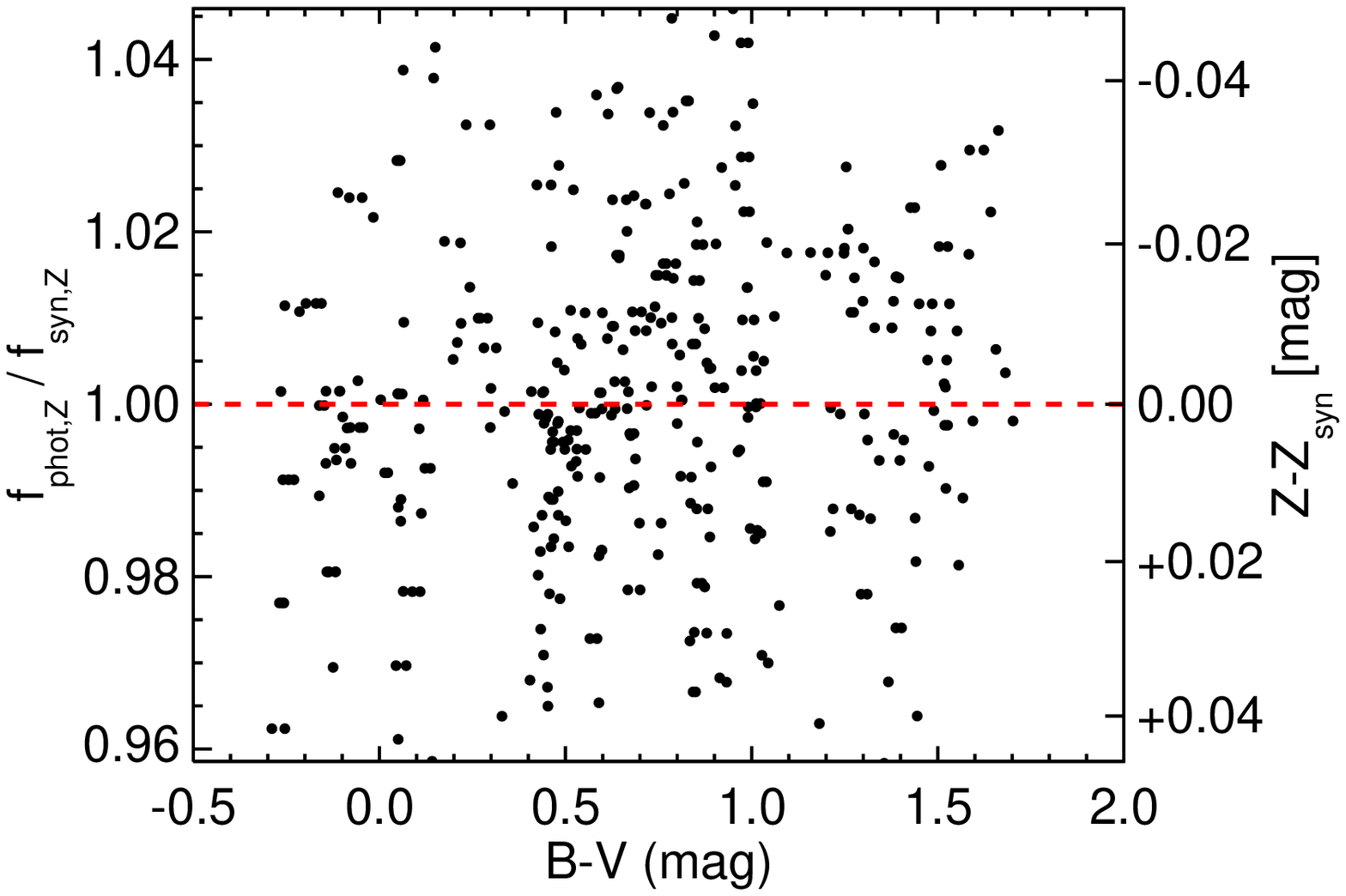} 
   \caption{Similar to Figure~\ref{fig:ratio1}, but using our revised filter profiles and zero points. Note that the Y-axis scales of the two plots are not the same, nor are they the same as those in Figure~\ref{fig:ratio1}. For $Z$ and $U$ respectively there are 3 and 8 outlier points outside the scale of the Figures. These are most likely variable stars, erroneous measurements, or measurements assigned to the wrong star as the scatter in repeat measurements with the same filter on the same star is typically only 0.02-0.04~mags. The red dashed line indicates agreement between photometric and spectroscopic fluxes. Consult Section~\ref{sec:newcal} for more information.}
   \label{fig:ratio2}
\end{figure*}

In Figure~\ref{fig:ratio2} we show $f_{phot,x}/f_{syn,x}$ as a function of $B-V$ for two sample filters using our revised curves and zero points. Inspection of analogous plots for all filters and systems indicates that our profiles are removing systematic trends in the data. 

\begin{figure}[tbp] 
   \centering
   \includegraphics[width=0.5\textwidth]{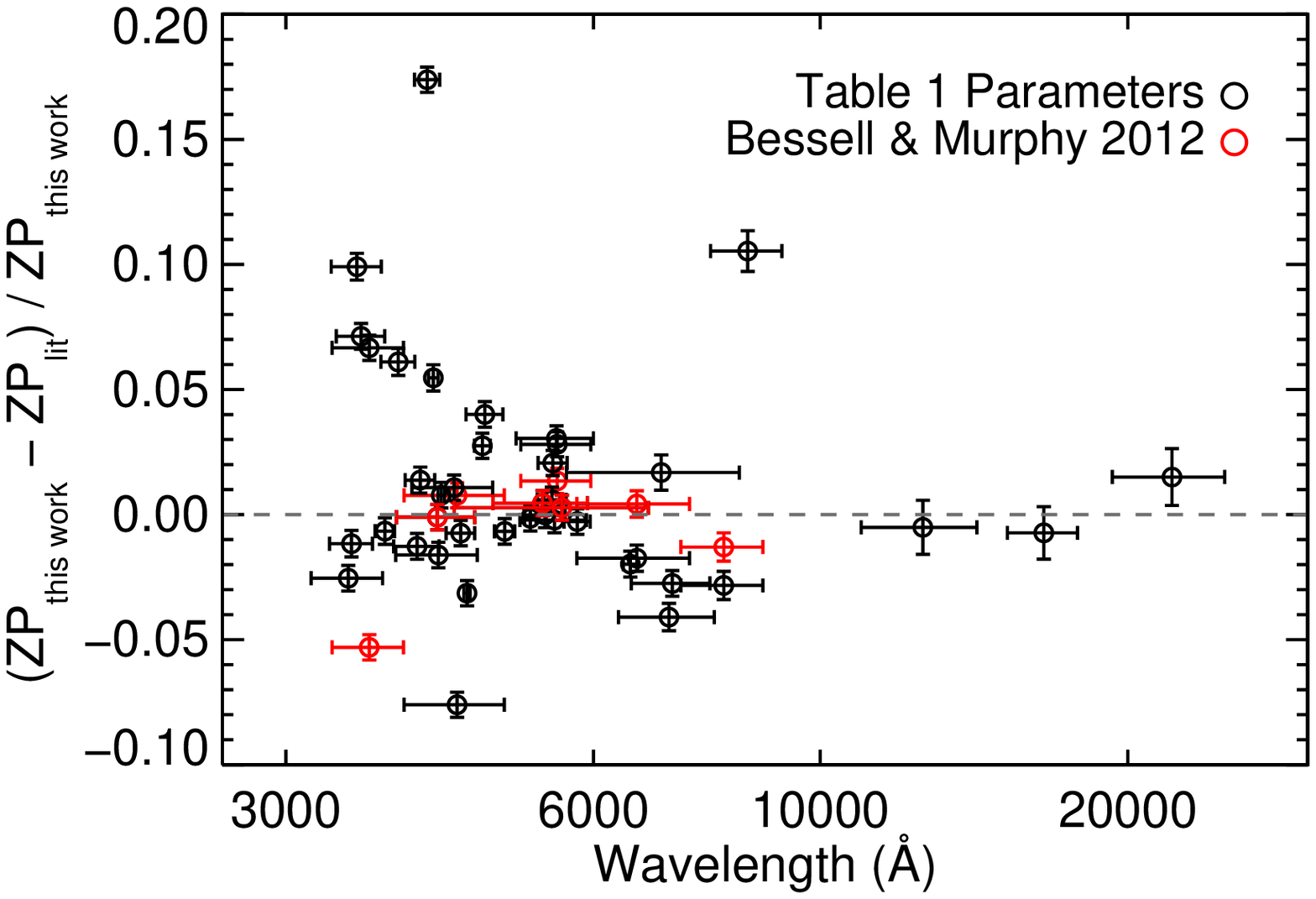} 
   \caption{Fractional difference between zero points derived in this work and those from Table~\ref{tab:phot}. Horizontal error bars indicate the \weff for the given filter. Vertical error bars show the estimated error in our own zero points. While many filters change by $<5\%$, five change by more than 10\%, including three off the scale of the Figure. Zero points from \citet{2012PASP..124..140B} are shown (in red) for reference. See Section~\ref{sec:results} for more details.}
   \label{fig:zp_diff}
\end{figure}

In Figure~\ref{fig:zp_diff} we show a comparison of the input (original) zero points to those derived from our MC analysis. Of the 42 filters tested, 32 change by $<5\%$, 6 change by 5--10\%, and 4 change by $>10\%$. Input zero points are meant as an initial guess for the MC, and are not necessarily the best available. Thus we also compare to the more recent analysis of \citet{2012PASP..124..140B} for Johnson $UBV$, Cousins $RI$, \hipp, and Tycho-2 photometry. Our zero points are within 2\% of those from \citet{2012PASP..124..140B}, with the exception of the Johnson $U$ filter, for which we disagree by 5\%. We suspect the difference is due to their use of the MILES library, which does not cover the full $U$ band.

\begin{figure}[tbp] 
   \centering
   \includegraphics[width=0.5\textwidth]{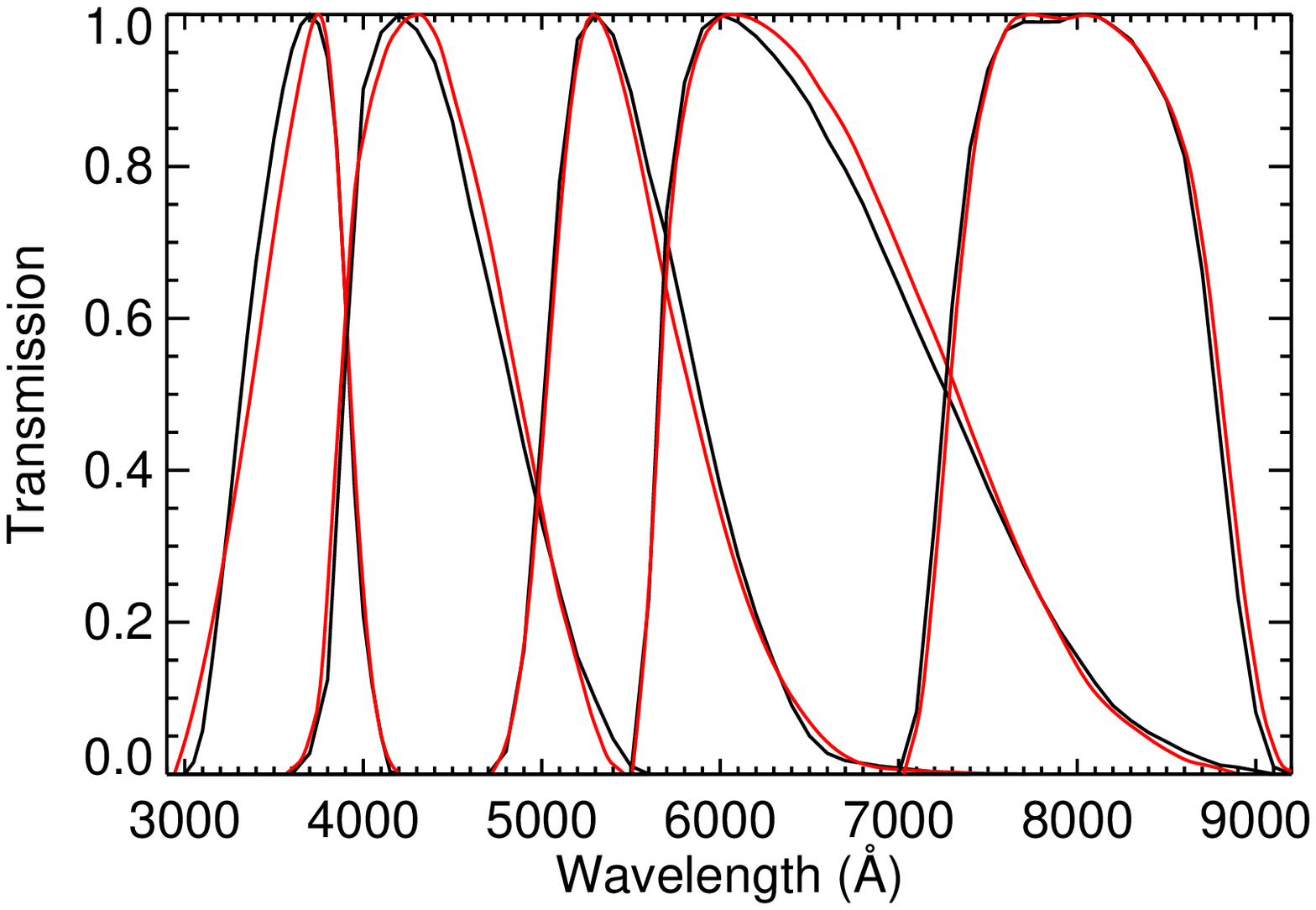} 
   \caption{Filter response curves derived from our analysis (red) compared to those from \citet{2012PASP..124..140B} for Johnson $UBV$ and Cousins $RI$ filters. Note all filters are shown as energy response curves.}
   \label{fig:comp_ubv}
\end{figure}

\begin{figure}[tbp] 
   \centering
   \includegraphics[width=0.5\textwidth]{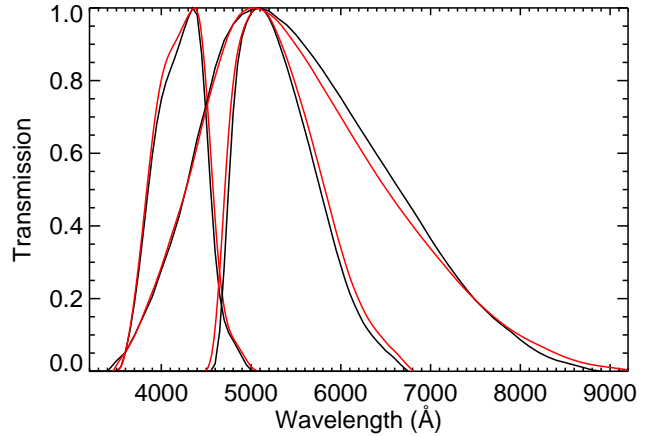} 
   \caption{Same as Figure~\ref{fig:comp_ubv} but for \hipp\ and Tycho-2 passbands. The broadest passband is the \hipp\ filter.}
   \label{fig:comp_tychip}
\end{figure}

\begin{figure}[tbp] 
   \centering
   \includegraphics[width=0.5\textwidth]{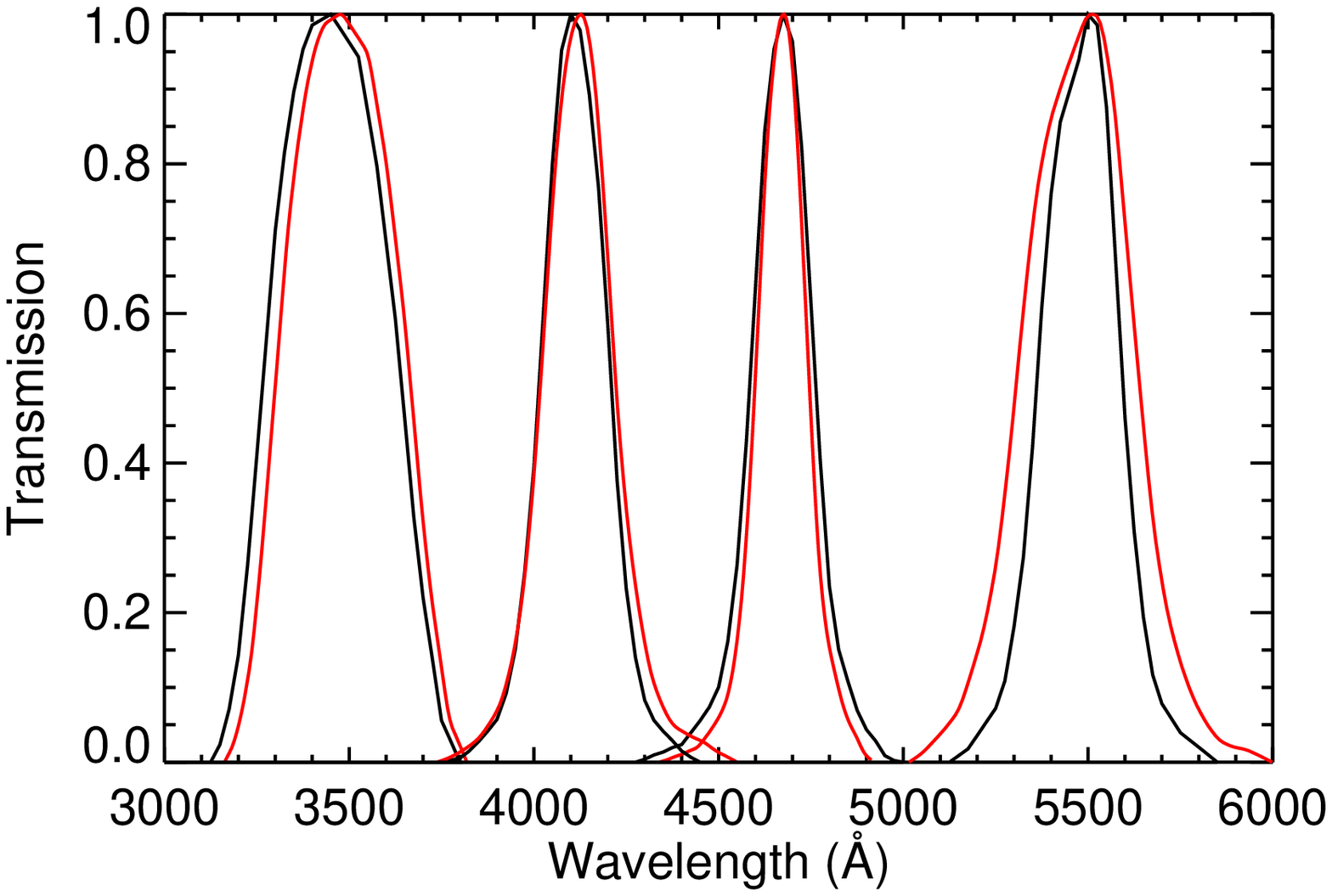} 
   \caption{Same as Figure~\ref{fig:comp_ubv} but for Str\"omgren ($ubvy$) passbands from \citet{2011PASP..123.1442B}. }
   \label{fig:comp_strom}
\end{figure}

In Figures~\ref{fig:comp_ubv}, \ref{fig:comp_tychip}, and \ref{fig:comp_strom} we compare our derived filter profiles to those from \citet{2011PASP..123.1442B} and \citet{2012PASP..124..140B}. The differences appear relatively small, but small differences can have a big impact for the bluest and reddest stars (see Figures~\ref{fig:ratio1} and \ref{fig:bessell_trend}). Note that zero point and filter profile changes are also correlated. If we adopt the input filter profiles we derive zero points which are generally closer to the originals, but this reintroduces color-terms.

\subsection{Understanding Errors}\label{sec:errors}
One important aspect of our analysis is to come up with an estimate for the error on the zero points. A simple method would be to calculate the standard error on the ratios of observed to synthetic fluxes ($f_{phot,x}/f_{syn,x}$). Because most bands have $\gg100$ points and the standard deviation of $f_{phot,x}/f_{syn,x}$ within a given filter tends to be small (1--3\%), this method would suggest zero point errors $\lesssim0.2\%$. This is unrealistically small because our analysis is dominated by systematic errors. In particular the largest source of error is errors in the spectrophotometric calibration of the NGSL or IRTF spectra. 

Flux calibration better than 1\% in the NIR is extraordinarily difficult because time-dependent changes in the sky (seeing, transmission, etc.) create wavelength-dependent flux losses that vary between target and standard observations. \citet{Cushing:2005lr} and \citet{Rayner:2009kx} mitigate for this by observing at the parallactic angle. By absolutely calibrating the library to 2MASS and comparing the residuals, \citet{Rayner:2009kx} estimate that the IRTF library has spectral slopes accurate to 1--3\% percent, although our own analysis suggests it is $\simeq1\%$, and much of the variation actually arises because of larger errors in 2MASS photometry for the brightest stars (non-linear part of the detector or saturation). Differences between 2MASS and the IRTF spectra are consistent with random errors, but we cannot rule out (systematic) slope errors of 1\%. To be conservative we add 1\% errors in quadrature with the standard errors calculated for zero points on $JHK$ photometry.

Space-based spectra should be able to achieve extremely accurate relative flux calibrations, since the major source of error for ground-based observations is removal of time-dependent noise from the sky \citep{Young:1991lr,Mann:2011qy}. However, the instrument and telescope can also introduce slope changes. In fact, an earlier version of the NGSL had significant errors in the spectral shape due to imperfect centering of the target on the slit. This was later calibrated out in post-processing, but highlights the difficulties in precisely flux calibrating spectra.

\begin{figure}[tbp] 
   \centering
   \includegraphics[width=0.5\textwidth]{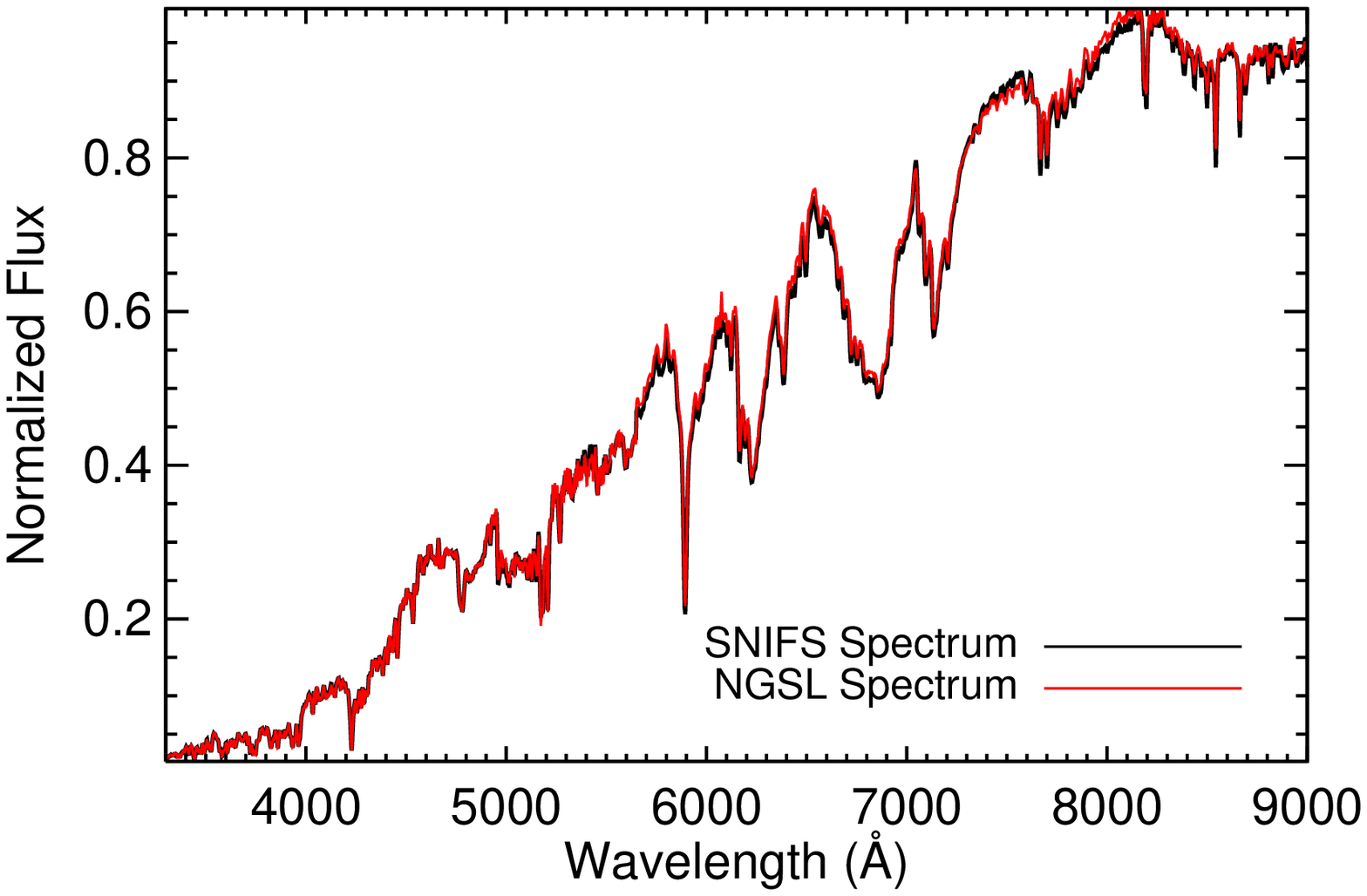} 
   \caption{Ground-based optical spectrum (black) taken from \cite{2013ApJ...779..188M} compared to a STIS spectrum (red) from NGSL of the star HD 1326 (GJ 15A). Both spectra have been normalized to their peak value. The differences in slope are  $\simeq1\%$, which is consistent with expected errors from SNIFS. See Section~\ref{sec:errors} for more information.}
   \label{fig:snifs}
\end{figure}

To test the level of noise expected from shape errors we compare NGSL spectra with well flux-calibrated ground-based spectra from \citet{2013ApJ...779..188M} for 10 overlapping stars (see Figure~\ref{fig:snifs} for an example). Slope differences are $<2\%$ in all cases, and on average are 0.9\%. \citet{2013ApJ...779..188M} report slope errors of 0.5--1\% in their spectra based on repeat observations. This suggests most of the difference between the ground- and space-based spectra can be explained by errors in the ground-based data. With this small comparison sample we cannot rule out slope errors as large as 0.5\%, so we conservatively add this to the standard errors on the zero points for all optical photometry. Final zero point errors are listed in Table~\ref{tab:final} for all systems included in our analysis.

\begin{figure}[tbp] 
   \centering
   \includegraphics[width=0.5\textwidth]{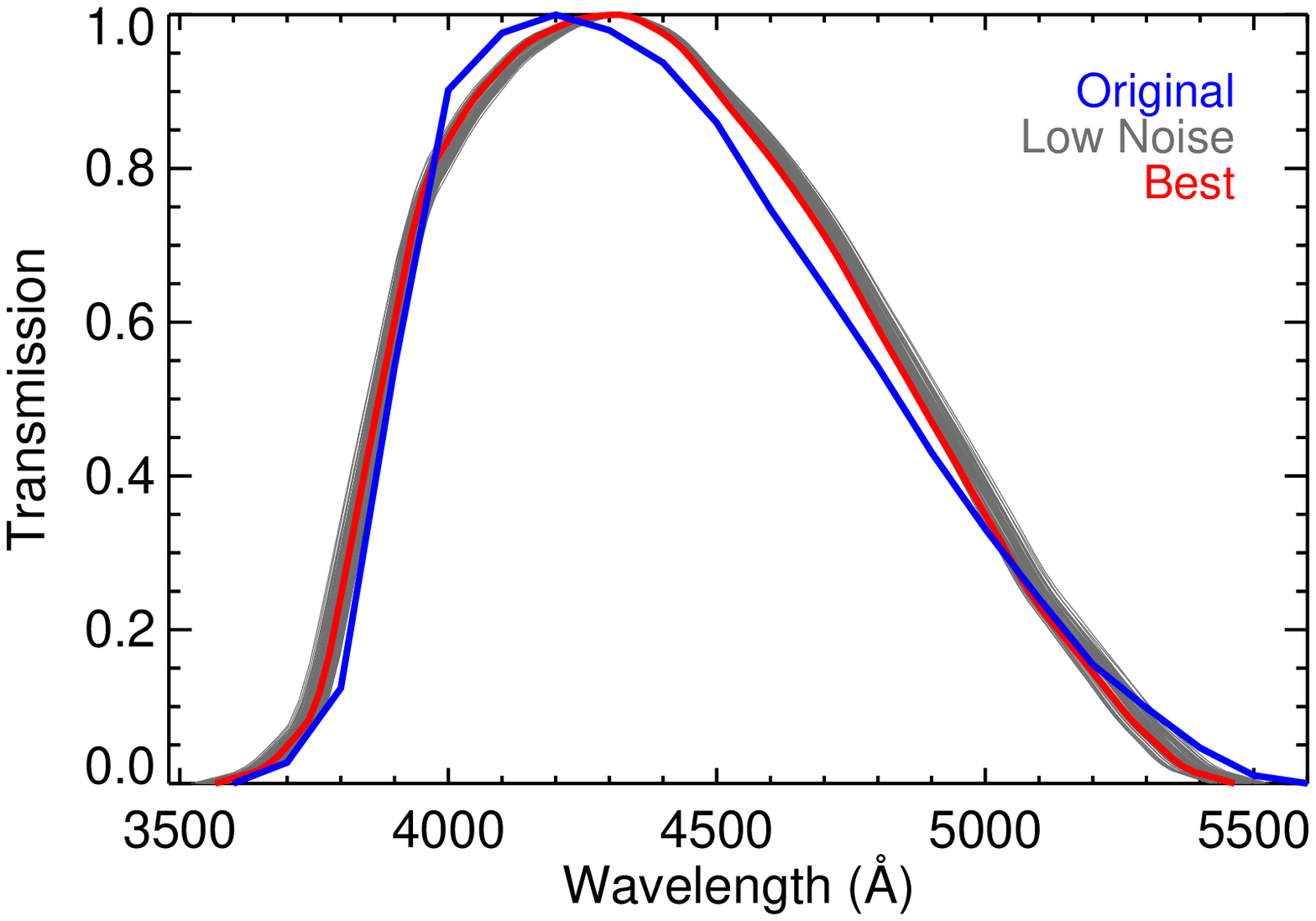} 
   \caption{Filter profile for the Johnson $B$ filter with original denoted in blue, and the best (adopted) profile shown in red. All filter profiles that show no correlation between $f_{phot,x}/f_{syn,x}$ and $B-V$ and have an RMS in $f_{phot,x}/f_{syn,x}$ less than double the minimum RMS (ones we consider acceptable solutions) that came out of our MC analysis are shown in grey. The original filter has been converted to energy response (as opposed to photon counting) for consistency with our analysis. See Section~\ref{sec:errors} for more information.}
   \label{fig:filter_range}
\end{figure}

For any given filter our analysis produces a large number of acceptable filter profiles, i.e. profiles which have no statistically significant color terms and an RMS in $f_{phot,x}/f_{syn,x}$ within a factor of two of the smallest RMS (we adopt the smallest RMS as the true filter). In Figure~\ref{fig:filter_range} we show an example distribution of all acceptable filter profiles that comes out of our MC analysis. For all filters the distribution is similarly tight, which suggests that the adopted profiles are fairly reliable and that we are getting close to the real system parameters. This also suggests that errors on our derived filter profiles are relatively small. 

\section{Summary and Discussion}\label{sec:conclusion}
In our effort to improve measurements of \fbol{} and, by extension \teff{} and/or $R_*$, we tested the zero points and passbands for 42 different filters across a range of  photometric systems. To this end we took advantage of the precise relative flux calibration of the NGSL and IRTF spectral libraries. By comparing flux calculated from the photometry to the corresponding flux in the spectrum we showed that errors in zero points and filter passbands can lead to flux errors of $>25\%$ for the bluest and reddest stars. 

To derive new passbands and zero points, we ran an MC analysis on each of the 42 filters in our sample. The MC found the filter parameters that produce the best agreement between spectroscopic and photometric fluxes free of correlations with the color of the star. For most bands the changes were relatively small (e.g., $<5\%$ in zero points), but some changed by $>10\%$. Changes in filter profiles were also generally small, but still had a very significant effect on stars with extreme colors, and hence are important to avoiding systematic errors.

We estimated errors in our zero points by considering possible shape (relative flux calibration) errors in the NGSL and IRTF spectra. Based on a comparison to spectra from other sources, we estimated that the relative flux calibration of the NGSL library is good to $\lesssim0.5\%$, although this may be a conservative estimate. For the IRTF spectra, the flux calibration is worse because of the complications involved in flux calibration NIR spectra from the ground. We expect that zero points for NIR filters are still good to $\simeq1\%$ for most bands. 

We also have estimated errors on the filter profiles by considering the range of reasonable filter profiles that come out of our MC analysis (Figure~\ref{fig:filter_range}). We report these errors alongside our derived filter profiles in the Appendix. However, readers are cautioned that these errors are highly correlated. For example, if we adopt transmission values lower than the best-fit value for the red end of the filter, then it's likely we will have to reduce the throughput in the blue in order to avoid introducing a color term. 

Our analysis has assumed that the true filter profile is a simple correction to the original profile. More complicated changes are more difficult to take into account because it makes parameter space too wide for the number of data points. For example, red leaks, which are common for blue filters, can easily give rise to color terms like those seen in this paper. However, most CCDs optimized for blue filters have low quantum efficiency in the red, and hence are only mildly affected by red leaks. Given that we are able to derive filter profiles free of color terms even for photometry taken from different sources, using different detectors and optical systems suggests that few, if any of these filters are subject to very significant red leaks. Future analysis with additional spectra will enable us to search using more complicated filter profiles and confirm this suggestion empirically.

The presented revisions (see Table~\ref{tab:final}) are derived based on data in our spectroscopic sample that span the range $-0.4\lesssim B-V \lesssim1.7$. Consequently, we caution the reader that there may be color-terms in the filter profiles that only become apparent at more extreme red and blue colors.

We note that our derived system sensitivities are not necessarily the true profile. It is certainly possible that a system sensitivity could produce no color terms and still not match the original standards, although this raises the issue of whether it is possible to define a true system in an absolute sense. Original system definitions are rarely determined with the precision we can achieve today. It is unlikely that CCDs and filters installed are perfect clones of the original system, and even if they were, filters/instruments are often changed and observations are corrected using imperfect transformations to put data back onto the standard systems. This data amalgamation makes it difficult to define a single system sensitivity that describes all the photometry. However, narrow range of system profiles that comes out of our MC analysis is promising, and suggestive that even if our parameters do not match the original system they are useful and accurate for calculating synthetic photometry and fluxes. 

Future efforts to calibrate NIR photometry will require spectra with better spectrophotometric calibration. The X-Shooter library \citep{2011JPhCS.328a2023C} offers a special advantage because it is build from optical and NIR spectra taken simultaneously. In this case optical photometry can be used to absolutely flux calibrate the spectrum, and then the NIR region can be used to independently test the NIR photometry. Unfortunately, the currently available version of the X-shooter library contains only optical data. Another advance will come with the expansion of the NGSL library (HST approved program ID 13776). This is expected to widen the color of the sample and provide an opportunity to revise/test the spectrophotometric calibration. 

One of the primary motivations of this study was to significantly improve on the precision and accuracy of derived bolometric fluxes for stars observed with long-baseline optical interferometry. The shifts in filter profiles are especially important for this, because color terms will lead to systematic differences in the absolute photometry. This will correspondingly shift the absolute flux calibration, and hence, the derived \fbol{} and \teff{} for any given star. 

\begin{figure}[t] 
   \centering
   \includegraphics[width=0.5\textwidth]{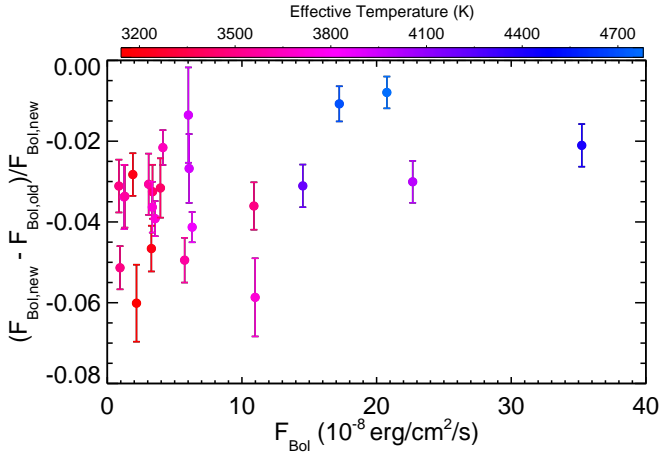} 
   \caption{Values of \fbol{} calculated using updated parameters from this work (new) compared to those calculated using the original input parameters (old). Errors are calculated from the scatter in the normalization and measurement errors in the spectrum.}
   \label{fig:fbol_comp}
\end{figure}

We highlight this point by recalculating \fbol{} for the K and M dwarfs from \citet{Boyajian2012} and \citet{2013ApJ...779..188M}, first using the original system parameters and then using the updated parameters from this work. We follow the same procedure to flux calibrate the spectra, i.e., use the median correction from all available photometry. As in \citet{2013ApJ...779..188M}, we extrapolate beyond the edges of the spectrum using a Rayleigh-Jeans law or Wien's approximation for the red and blue end, respectively. The \fbol{} is then calculated from the numerical integral of the entire spectrum. In Figure~\ref{fig:fbol_comp} we show the comparison of \fbol{} using old and new photometric system parameters. There is a clear systematic offset caused by the differences in absolute flux calibrations. This amounts to a median offset of 3\%, which is statistically significant for most stars, and is highly significant if we consider the full sample. Further, the offset is larger for the coolest (reddest) stars (3--7\%), emphasizing the importance of accurate photometric system parameters to reducing systematic errors.

\acknowledgments
The authors thank the referee, Ralph Bohlin, for his extremely quick and useful review of this work. We also thank Megan Ansdell, Eric Gaidos, and Chao-Ling Hung for their useful comments on this manuscript. We give additional to thanks Simon Murphy, Tabetha Boyajian, and Gerard van Belle for their useful discussions at Cool Stars 18. This work was supported by the Harlan J. Smith Fellowship from the University of Texas at Austin to AWM. 

This research has made use of the SIMBAD database, operated at CDS, Strasbourg, France. This work is based on observations made with the NASA/ESA Hubble Space Telescope, obtained at the Space Telescope Science Institute, which is operated by the Association of Universities for Research in Astronomy, Inc., under NASA contract NAS 5-26555. These observations are associated with GO program 11652.

{\it Facility:} \facility{IRTF}

\bibliography{$HOME/dropbox/fullbiblio.bib}

\clearpage

\appendix{}
Here we provide our final filter profiles. Electronic versions are available with this submission.

\begin{deluxetable}{l l | l l | l l | l l }
\tablewidth{0pt}
\tabletypesize{\scriptsize}
\renewcommand{\tabcolsep}{0.1cm}
\tablecaption{Energy Response Functions, $S_x(\lambda)$, for Johnson}
\tablehead{
\multicolumn{2}{c}{$U$} &\multicolumn{2}{c}{$B$} &\multicolumn{2}{c}{$V$} &\multicolumn{2}{c}{$R$} \\
\colhead{$\lambda$} & \colhead{Trans} &\colhead{$\lambda$} & \colhead{Trans} &\colhead{$\lambda$} & \colhead{Trans} &\colhead{$\lambda$} & \colhead{Trans} \\
\colhead{\AA} & \colhead{} &\colhead{\AA} & \colhead{} &\colhead{\AA} & \colhead{} &\colhead{\AA} & \colhead{} } 
\startdata
  2942.0 &  0.000 &   3566.0 &  0.000 &   4718.2 &  0.000 &   5945.3 &  0.000\\
  2959.1 &  0.012 &   3591.6 &  0.005 &   4758.6 &  0.017 &   5990.7 &  0.028\\
  2976.2 &  0.024 &   3617.3 &  0.010 &   4799.1 &  0.041 &   6036.0 &  0.061\\
  2993.3 &  0.037 &   3642.9 &  0.016 &   4839.5 &  0.078 &   6081.3 &  0.107\\
  3010.4 &  0.051 &   3668.6 &  0.027 &   4880.0 &  0.134 &   6126.7 &  0.179\\
  3027.5 &  0.066 &   3694.2 &  0.044 &   4920.4 &  0.211 &   6172.0 &  0.275\\
  3044.6 &  0.082 &   3719.9 &  0.064 &   4960.8 &  0.311 &   6217.3 &  0.381\\
  3061.7 &  0.098 &   3745.5 &  0.092 &   5001.3 &  0.429 &   6262.6 &  0.479\\
\enddata
\label{tab:sample}
\end{deluxetable}
 
\begin{deluxetable}{l l }
\tablewidth{0pt}
\tabletypesize{\scriptsize}
\renewcommand{\tabcolsep}{0.1cm}
\tablecaption{Energy Response Functions, $S_x(\lambda)$, for Cape}
\tablehead{
\multicolumn{2}{c}{$U$} \\
\colhead{$\lambda$} & \colhead{Trans} \\
\colhead{\AA} & \colhead{} } 
\startdata
  3190.4 &  0.000\\
  3215.4 &  0.003\\
  3240.3 &  0.006\\
  3265.3 &  0.009\\
  3290.3 &  0.012\\
  3315.3 &  0.014\\
  3340.3 &  0.017\\
  3365.3 &  0.021\\
  3390.2 &  0.030\\
  3415.2 &  0.043\\
  3440.2 &  0.065\\
  3465.2 &  0.098\\
  3490.2 &  0.139\\
  3515.2 &  0.188\\
  3540.1 &  0.246\\
  3565.1 &  0.309\\
  3590.1 &  0.379\\
  3615.1 &  0.452\\
  3640.1 &  0.527\\
  3665.1 &  0.604\\
  3690.0 &  0.679\\
  3715.0 &  0.751\\
  3740.0 &  0.820\\
  3765.0 &  0.880\\
  3790.0 &  0.931\\
  3814.9 &  0.969\\
  3839.9 &  0.991\\
  3864.9 &  1.000\\
  3889.9 &  0.986\\
  3914.9 &  0.950\\
  3939.9 &  0.897\\
  3964.8 &  0.820\\
  3989.8 &  0.726\\
  4014.8 &  0.626\\
  4039.8 &  0.522\\
  4064.8 &  0.422\\
  4089.8 &  0.339\\
  4114.7 &  0.275\\
  4139.7 &  0.222\\
  4164.7 &  0.186\\
  4189.7 &  0.163\\
  4214.7 &  0.143\\
  4239.7 &  0.129\\
  4264.6 &  0.118\\
  4289.6 &  0.109\\
  4314.6 &  0.102\\
  4339.6 &  0.096\\
  4364.6 &  0.091\\
  4389.6 &  0.088\\
  4414.5 &  0.086\\
  4439.5 &  0.085\\
  4464.5 &  0.086\\
  4489.5 &  0.088\\
  4514.5 &  0.092\\
  4539.5 &  0.097\\
  4564.4 &  0.101\\
  4589.4 &  0.106\\
  4614.4 &  0.110\\
  4639.4 &  0.113\\
  4664.4 &  0.114\\
  4689.3 &  0.113\\
  4714.3 &  0.110\\
  4739.3 &  0.105\\
  4764.3 &  0.098\\
  4789.3 &  0.090\\
  4814.3 &  0.080\\
  4839.2 &  0.070\\
  4864.2 &  0.059\\
  4889.2 &  0.048\\
  4914.2 &  0.039\\
  4939.2 &  0.030\\
  4964.2 &  0.022\\
  4989.1 &  0.014\\
  5014.1 &  0.007
\enddata
\label{tab:sample}
\end{deluxetable}
 
\begin{deluxetable}{l l | l l | l l | l l | l l | l l }
\tablewidth{0pt}
\tabletypesize{\scriptsize}
\renewcommand{\tabcolsep}{0.1cm}
\tablecaption{Energy Response Functions, $S_x(\lambda)$, for DDO}
\tablehead{
\multicolumn{2}{c}{$m35$} &\multicolumn{2}{c}{$m38$} &\multicolumn{2}{c}{$m41$} &\multicolumn{2}{c}{$m42$} &\multicolumn{2}{c}{$m45$} &\multicolumn{2}{c}{$m48$} \\
\colhead{$\lambda$} & \colhead{Trans} &\colhead{$\lambda$} & \colhead{Trans} &\colhead{$\lambda$} & \colhead{Trans} &\colhead{$\lambda$} & \colhead{Trans} &\colhead{$\lambda$} & \colhead{Trans} &\colhead{$\lambda$} & \colhead{Trans} \\
\colhead{\AA} & \colhead{} &\colhead{\AA} & \colhead{} &\colhead{\AA} & \colhead{} &\colhead{\AA} & \colhead{} &\colhead{\AA} & \colhead{} &\colhead{\AA} & \colhead{} } 
\startdata
  3166.1 &  0.000 &   3605.5 &  0.000 &   4094.0 &  0.000 &   4199.0 &  0.000 &   4454.2 &  0.000 &   4695.1 &  0.000\\
  3176.1 &  0.004 &   3613.0 &  0.011 &   4096.4 &  0.002 &   4200.7 &  0.003 &   4455.9 &  0.004 &   4700.8 &  0.008\\
  3186.1 &  0.008 &   3620.6 &  0.024 &   4098.8 &  0.004 &   4202.3 &  0.005 &   4457.6 &  0.008 &   4706.4 &  0.013\\
  3196.1 &  0.017 &   3628.1 &  0.041 &   4101.2 &  0.006 &   4204.0 &  0.010 &   4459.2 &  0.013 &   4712.1 &  0.016\\
  3206.1 &  0.030 &   3635.6 &  0.063 &   4103.6 &  0.010 &   4205.7 &  0.016 &   4460.9 &  0.019 &   4717.7 &  0.019\\
  3216.1 &  0.047 &   3643.2 &  0.090 &   4106.0 &  0.015 &   4207.3 &  0.023 &   4462.6 &  0.026 &   4723.4 &  0.023\\
  3226.1 &  0.071 &   3650.7 &  0.121 &   4108.4 &  0.022 &   4209.0 &  0.030 &   4464.2 &  0.035 &   4729.0 &  0.028\\
  3236.1 &  0.102 &   3658.2 &  0.158 &   4110.8 &  0.030 &   4210.6 &  0.037 &   4465.9 &  0.048 &   4734.7 &  0.035\\
\enddata
\label{tab:sample}
\end{deluxetable}
 
\begin{deluxetable}{l l | l l }
\tablewidth{0pt}
\tabletypesize{\scriptsize}
\renewcommand{\tabcolsep}{0.1cm}
\tablecaption{Energy Response Functions, $S_x(\lambda)$, for Eggen}
\tablehead{
\multicolumn{2}{c}{$R$} &\multicolumn{2}{c}{$I$} \\
\colhead{$\lambda$} & \colhead{Trans} &\colhead{$\lambda$} & \colhead{Trans} \\
\colhead{\AA} & \colhead{} &\colhead{\AA} & \colhead{} } 
\startdata
  5010.7 &  0.000 &   7309.7 &  0.000\\
  5071.3 &  0.001 &   7349.2 &  0.004\\
  5131.9 &  0.004 &   7388.6 &  0.009\\
  5192.5 &  0.012 &   7428.0 &  0.019\\
  5253.1 &  0.026 &   7467.4 &  0.034\\
  5313.7 &  0.055 &   7506.8 &  0.055\\
  5374.3 &  0.116 &   7546.3 &  0.083\\
  5434.9 &  0.221 &   7585.7 &  0.118\\
\enddata
\label{tab:sample}
\end{deluxetable}
 
\begin{deluxetable}{l l | l l | l l | l l | l l | l l | l l }
\tablewidth{0pt}
\tabletypesize{\scriptsize}
\renewcommand{\tabcolsep}{0.1cm}
\tablecaption{Energy Response Functions, $S_x(\lambda)$, for Geneva}
\tablehead{
\multicolumn{2}{c}{$U$} &\multicolumn{2}{c}{$B1$} &\multicolumn{2}{c}{$B$} &\multicolumn{2}{c}{$B2$} &\multicolumn{2}{c}{$V1$} &\multicolumn{2}{c}{$V$} &\multicolumn{2}{c}{$G$} \\
\colhead{$\lambda$} & \colhead{Trans} &\colhead{$\lambda$} & \colhead{Trans} &\colhead{$\lambda$} & \colhead{Trans} &\colhead{$\lambda$} & \colhead{Trans} &\colhead{$\lambda$} & \colhead{Trans} &\colhead{$\lambda$} & \colhead{Trans} &\colhead{$\lambda$} & \colhead{Trans} \\
\colhead{\AA} & \colhead{} &\colhead{\AA} & \colhead{} &\colhead{\AA} & \colhead{} &\colhead{\AA} & \colhead{} &\colhead{\AA} & \colhead{} &\colhead{\AA} & \colhead{} &\colhead{\AA} & \colhead{} } 
\startdata
  2909.4 &  0.000 &   3533.3 &  0.000 &   3580.5 &  0.000 &   4115.7 &  0.000 &   4972.4 &  0.000 &   4838.4 &  0.000 &   5498.3 &  0.000\\
  2923.5 &  0.018 &   3550.3 &  0.001 &   3600.8 &  0.018 &   4125.1 &  0.004 &   4991.0 &  0.058 &   4864.2 &  0.023 &   5508.8 &  0.029\\
  2937.6 &  0.036 &   3567.3 &  0.002 &   3621.2 &  0.036 &   4134.4 &  0.008 &   5009.6 &  0.116 &   4890.0 &  0.046 &   5519.4 &  0.059\\
  2951.7 &  0.054 &   3584.4 &  0.003 &   3641.6 &  0.055 &   4143.8 &  0.012 &   5028.2 &  0.175 &   4915.7 &  0.071 &   5530.0 &  0.090\\
  2965.9 &  0.072 &   3601.4 &  0.004 &   3661.9 &  0.076 &   4153.2 &  0.016 &   5046.8 &  0.233 &   4941.5 &  0.121 &   5540.5 &  0.123\\
  2980.0 &  0.090 &   3618.5 &  0.006 &   3682.3 &  0.109 &   4162.5 &  0.020 &   5065.3 &  0.300 &   4967.3 &  0.213 &   5551.1 &  0.160\\
  2994.1 &  0.108 &   3635.5 &  0.016 &   3702.6 &  0.157 &   4171.9 &  0.025 &   5083.9 &  0.389 &   4993.1 &  0.333 &   5561.6 &  0.211\\
  3008.2 &  0.127 &   3652.6 &  0.038 &   3723.0 &  0.210 &   4181.2 &  0.032 &   5102.5 &  0.500 &   5018.9 &  0.457 &   5572.2 &  0.285\\
\enddata
\label{tab:sample}
\end{deluxetable}
 
\begin{deluxetable}{l l | l l | l l | l l }
\tablewidth{0pt}
\tabletypesize{\scriptsize}
\renewcommand{\tabcolsep}{0.1cm}
\tablecaption{Energy Response Functions, $S_x(\lambda)$, for Stromgren}
\tablehead{
\multicolumn{2}{c}{$u$} &\multicolumn{2}{c}{$b$} &\multicolumn{2}{c}{$v$} &\multicolumn{2}{c}{$y$} \\
\colhead{$\lambda$} & \colhead{Trans} &\colhead{$\lambda$} & \colhead{Trans} &\colhead{$\lambda$} & \colhead{Trans} &\colhead{$\lambda$} & \colhead{Trans} \\
\colhead{\AA} & \colhead{} &\colhead{\AA} & \colhead{} &\colhead{\AA} & \colhead{} &\colhead{\AA} & \colhead{} } 
\startdata
  3162.5 &  0.000 &   4339.1 &  0.000 &   3737.4 &  0.000 &   5015.6 &  0.000\\
  3171.4 &  0.008 &   4346.8 &  0.001 &   3748.3 &  0.002 &   5028.9 &  0.004\\
  3180.2 &  0.017 &   4354.6 &  0.002 &   3759.3 &  0.004 &   5042.2 &  0.009\\
  3189.1 &  0.029 &   4362.4 &  0.004 &   3770.3 &  0.006 &   5055.4 &  0.014\\
  3198.0 &  0.045 &   4370.1 &  0.005 &   3781.3 &  0.009 &   5068.7 &  0.021\\
  3206.8 &  0.063 &   4377.9 &  0.007 &   3792.3 &  0.011 &   5082.0 &  0.028\\
  3215.7 &  0.084 &   4385.7 &  0.008 &   3803.2 &  0.014 &   5095.3 &  0.035\\
  3224.6 &  0.109 &   4393.4 &  0.009 &   3814.2 &  0.018 &   5108.5 &  0.043\\
\enddata
\label{tab:sample}
\end{deluxetable}
 
\begin{deluxetable}{l l | l l | l l | l l | l l | l l | l l }
\tablewidth{0pt}
\tabletypesize{\scriptsize}
\renewcommand{\tabcolsep}{0.1cm}
\tablecaption{Energy Response Functions, $S_x(\lambda)$, for Vilnius}
\tablehead{
\multicolumn{2}{c}{$U$} &\multicolumn{2}{c}{$P$} &\multicolumn{2}{c}{$X$} &\multicolumn{2}{c}{$Y$} &\multicolumn{2}{c}{$Z$} &\multicolumn{2}{c}{$V$} &\multicolumn{2}{c}{$S$} \\
\colhead{$\lambda$} & \colhead{Trans} &\colhead{$\lambda$} & \colhead{Trans} &\colhead{$\lambda$} & \colhead{Trans} &\colhead{$\lambda$} & \colhead{Trans} &\colhead{$\lambda$} & \colhead{Trans} &\colhead{$\lambda$} & \colhead{Trans} &\colhead{$\lambda$} & \colhead{Trans} \\
\colhead{\AA} & \colhead{} &\colhead{\AA} & \colhead{} &\colhead{\AA} & \colhead{} &\colhead{\AA} & \colhead{} &\colhead{\AA} & \colhead{} &\colhead{\AA} & \colhead{} &\colhead{\AA} & \colhead{} } 
\startdata
  3047.3 &  0.000 &   3533.8 &  0.000 &   3674.0 &  0.000 &   4346.0 &  0.000 &   4873.4 &  0.000 &   4985.5 &  0.000 &   5785.1 &  0.000\\
  3057.6 &  0.002 &   3539.5 &  0.003 &   3686.7 &  0.005 &   4360.0 &  0.022 &   4882.0 &  0.001 &   5000.0 &  0.001 &   5809.5 &  0.004\\
  3067.9 &  0.003 &   3545.1 &  0.006 &   3699.5 &  0.009 &   4373.9 &  0.044 &   4890.6 &  0.002 &   5014.5 &  0.002 &   5833.8 &  0.008\\
  3078.2 &  0.005 &   3550.8 &  0.009 &   3712.2 &  0.014 &   4387.9 &  0.067 &   4899.2 &  0.003 &   5029.0 &  0.003 &   5858.2 &  0.011\\
  3088.4 &  0.008 &   3556.5 &  0.013 &   3725.0 &  0.020 &   4401.9 &  0.101 &   4907.8 &  0.004 &   5043.5 &  0.005 &   5882.5 &  0.016\\
  3098.7 &  0.012 &   3562.1 &  0.016 &   3737.7 &  0.029 &   4415.9 &  0.157 &   4916.4 &  0.004 &   5058.0 &  0.009 &   5906.9 &  0.022\\
  3109.0 &  0.018 &   3567.8 &  0.022 &   3750.5 &  0.039 &   4429.8 &  0.236 &   4925.0 &  0.005 &   5072.5 &  0.015 &   5931.3 &  0.030\\
  3119.3 &  0.025 &   3573.5 &  0.030 &   3763.3 &  0.050 &   4443.8 &  0.325 &   4933.6 &  0.005 &   5087.0 &  0.021 &   5955.6 &  0.037\\
\enddata
\label{tab:sample}
\end{deluxetable}
 
\begin{deluxetable}{l l | l l | l l | l l }
\tablewidth{0pt}
\tabletypesize{\scriptsize}
\renewcommand{\tabcolsep}{0.1cm}
\tablecaption{Energy Response Functions, $S_x(\lambda)$, for WBVR}
\tablehead{
\multicolumn{2}{c}{$W$} &\multicolumn{2}{c}{$B$} &\multicolumn{2}{c}{$V$} &\multicolumn{2}{c}{$R$} \\
\colhead{$\lambda$} & \colhead{Trans} &\colhead{$\lambda$} & \colhead{Trans} &\colhead{$\lambda$} & \colhead{Trans} &\colhead{$\lambda$} & \colhead{Trans} \\
\colhead{\AA} & \colhead{} &\colhead{\AA} & \colhead{} &\colhead{\AA} & \colhead{} &\colhead{\AA} & \colhead{} } 
\startdata
  2908.2 &  0.000 &   3486.0 &  0.000 &   4619.9 &  0.000 &   6039.9 &  0.000\\
  2923.2 &  0.000 &   3510.9 &  0.000 &   4650.4 &  0.001 &   6084.0 &  0.000\\
  2938.3 &  0.000 &   3535.8 &  0.000 &   4680.9 &  0.002 &   6128.1 &  0.002\\
  2953.3 &  0.001 &   3560.6 &  0.000 &   4711.4 &  0.004 &   6172.2 &  0.008\\
  2968.4 &  0.001 &   3585.5 &  0.001 &   4741.9 &  0.011 &   6216.3 &  0.037\\
  2983.4 &  0.002 &   3610.4 &  0.005 &   4772.4 &  0.023 &   6260.4 &  0.102\\
  2998.5 &  0.004 &   3635.2 &  0.013 &   4802.9 &  0.043 &   6304.4 &  0.222\\
  3013.5 &  0.006 &   3660.1 &  0.030 &   4833.4 &  0.079 &   6348.5 &  0.389\\
\enddata
\label{tab:sample}
\end{deluxetable}
 
\begin{deluxetable}{l l | l l }
\tablewidth{0pt}
\tabletypesize{\scriptsize}
\renewcommand{\tabcolsep}{0.1cm}
\tablecaption{Energy Response Functions, $S_x(\lambda)$, for Cousins}
\tablehead{
\multicolumn{2}{c}{$R$} &\multicolumn{2}{c}{$I$} \\
\colhead{$\lambda$} & \colhead{Trans} &\colhead{$\lambda$} & \colhead{Trans} \\
\colhead{\AA} & \colhead{} &\colhead{\AA} & \colhead{} } 
\startdata
  5507.0 &  0.000 &   7028.9 &  0.000\\
  5553.0 &  0.106 &   7058.8 &  0.025\\
  5598.9 &  0.241 &   7088.7 &  0.051\\
  5644.8 &  0.431 &   7118.6 &  0.083\\
  5690.7 &  0.631 &   7148.5 &  0.138\\
  5736.6 &  0.769 &   7178.4 &  0.210\\
  5782.5 &  0.853 &   7208.3 &  0.285\\
  5828.4 &  0.910 &   7238.2 &  0.366\\
\enddata
\label{tab:sample}
\end{deluxetable}
 
\begin{deluxetable}{l l | l l }
\tablewidth{0pt}
\tabletypesize{\scriptsize}
\renewcommand{\tabcolsep}{0.1cm}
\tablecaption{Energy Response Functions, $S_x(\lambda)$, for Tycho}
\tablehead{
\multicolumn{2}{c}{$Bt$} &\multicolumn{2}{c}{$Vt$} \\
\colhead{$\lambda$} & \colhead{Trans} &\colhead{$\lambda$} & \colhead{Trans} \\
\colhead{\AA} & \colhead{} &\colhead{\AA} & \colhead{} } 
\startdata
  3511.4 &  0.000 &   4494.7 &  0.000\\
  3532.8 &  0.007 &   4525.9 &  0.014\\
  3554.2 &  0.015 &   4557.2 &  0.042\\
  3575.5 &  0.029 &   4588.5 &  0.094\\
  3596.9 &  0.050 &   4619.7 &  0.175\\
  3618.3 &  0.074 &   4651.0 &  0.280\\
  3639.7 &  0.102 &   4682.3 &  0.399\\
  3661.1 &  0.134 &   4713.6 &  0.523\\
\enddata
\label{tab:sample}
\end{deluxetable}
 
\begin{deluxetable}{l l }
\tablewidth{0pt}
\tabletypesize{\scriptsize}
\renewcommand{\tabcolsep}{0.1cm}
\tablecaption{Energy Response Functions, $S_x(\lambda)$, for Hipparcos}
\tablehead{
\multicolumn{2}{c}{$Hp$} \\
\colhead{$\lambda$} & \colhead{Trans} \\
\colhead{\AA} & \colhead{} } 
\startdata
  3460.5 &  0.000\\
  3540.3 &  0.032\\
  3620.1 &  0.061\\
  3699.9 &  0.097\\
  3779.6 &  0.143\\
  3859.4 &  0.193\\
  3939.2 &  0.244\\
  4018.9 &  0.301\\
  4098.7 &  0.362\\
  4178.5 &  0.423\\
  4258.2 &  0.487\\
  4338.0 &  0.558\\
  4417.8 &  0.639\\
  4497.5 &  0.719\\
  4577.3 &  0.792\\
  4657.1 &  0.858\\
  4736.8 &  0.918\\
  4816.6 &  0.961\\
  4896.4 &  0.986\\
  4976.1 &  0.998\\
  5055.9 &  0.999\\
  5135.7 &  0.993\\
  5215.4 &  0.981\\
  5295.2 &  0.964\\
  5375.0 &  0.940\\
  5454.7 &  0.913\\
  5534.5 &  0.885\\
  5614.3 &  0.856\\
  5694.0 &  0.825\\
  5773.8 &  0.792\\
  5853.6 &  0.760\\
  5933.3 &  0.728\\
  6013.1 &  0.695\\
  6092.9 &  0.662\\
  6172.6 &  0.630\\
  6252.4 &  0.598\\
  6332.2 &  0.566\\
  6411.9 &  0.536\\
  6491.7 &  0.506\\
  6571.5 &  0.478\\
  6651.2 &  0.450\\
  6731.0 &  0.423\\
  6810.8 &  0.397\\
  6890.5 &  0.371\\
  6970.3 &  0.347\\
  7050.1 &  0.322\\
  7129.8 &  0.299\\
  7209.6 &  0.275\\
  7289.4 &  0.252\\
  7369.2 &  0.229\\
  7448.9 &  0.209\\
  7528.7 &  0.189\\
  7608.5 &  0.171\\
  7688.2 &  0.153\\
  7768.0 &  0.137\\
  7847.8 &  0.122\\
  7927.5 &  0.109\\
  8007.3 &  0.097\\
  8087.1 &  0.086\\
  8166.8 &  0.076\\
  8246.6 &  0.067\\
  8326.4 &  0.058\\
  8406.1 &  0.048\\
  8485.9 &  0.040\\
  8565.7 &  0.033\\
  8645.4 &  0.027\\
  8725.2 &  0.022\\
  8805.0 &  0.018\\
  8884.7 &  0.015\\
  8964.5 &  0.012\\
  9044.3 &  0.009\\
  9124.0 &  0.006\\
  9203.8 &  0.004\\
  9283.6 &  0.002
\enddata
\label{tab:sample}
\end{deluxetable}

\end{document}